\documentclass[onecolumn,amsmath,showpacs,nofootinbib,12pt]{revtex4-1}
\usepackage{graphicx}
\usepackage{dcolumn}
\usepackage{bm}
\usepackage{color}
\begin{document}
\newcommand{\hs}{\hspace*{0.5cm}}
\newcommand{\vs}{\vspace*{0.5cm}}
\newcommand{\be}{\begin{equation}}
\newcommand{\ee}{\end{equation}}
\newcommand{\bea}{\begin{eqnarray}}
\newcommand{\eea}{\end{eqnarray}}
\newcommand{\ben}{\begin{enumerate}}
\newcommand{\een}{\end{enumerate}}
\newcommand{\bde}{\begin{widetext}}
\newcommand{\ede}{\end{widetext}}
\newcommand{\nn}{\nonumber}
\newcommand{\crn}{\nonumber \\}
\newcommand{\Tr}{\mathrm{Tr}}
\newcommand{\non}{\nonumber}
\newcommand{\noi}{\noindent}
\newcommand{\al}{\alpha}
\newcommand{\la}{\lambda}
\newcommand{\bet}{\beta}
\newcommand{\ga}{\gamma}
\newcommand{\va}{\varphi}
\newcommand{\om}{\omega}
\newcommand{\pa}{\partial}
\newcommand{\+}{\dagger}
\newcommand{\fr}{\frac}
\newcommand{\bc}{\begin{center}}
\newcommand{\ec}{\end{center}}
\newcommand{\Ga}{\Gamma}
\newcommand{\de}{\delta}
\newcommand{\De}{\Delta}
\newcommand{\ep}{\epsilon}
\newcommand{\varep}{\varepsilon}
\newcommand{\ka}{\kappa}
\newcommand{\La}{\Lambda}
\newcommand{\si}{\sigma}
\newcommand{\Si}{\Sigma}
\newcommand{\ta}{\tau}
\newcommand{\up}{\upsilon}
\newcommand{\Up}{\Upsilon}
\newcommand{\ze}{\zeta}
\newcommand{\ps}{\psi}
\newcommand{\Ps}{\Psi}
\newcommand{\ph}{\phi}
\newcommand{\vph}{\varphi}
\newcommand{\Ph}{\Phi}
\newcommand{\Om}{\Omega}
\newcommand{\AdrHEPC}{$^3$Phenikaa Institute for Advanced Study, Phenikaa University, Hanoi 100000, Vietnam\\
$^4$Faculty of Basic Science and Faculty of Materials Science and Engineering,\\ Phenikaa University, Hanoi 100000, Vietnam}

\title{Flavor changing in the flipped trinification}
\author{D. N. Dinh$^1$, D. T. Huong$^1$, N. T. Duy$^2$,\\ N. T. Nhuan$^2$, L. D. Thien$^2$, and Phung Van Dong$^{3,4,*}$}
\affiliation{$^1$Institute of Physics, Vietnam Academy of Science and Technology, 10 Dao Tan, Ba Dinh, Hanoi, Vietnam\\
$^2$Graduate University of Science and Technology, Vietnam Academy of Science and Technology, 18 Hoang Quoc Viet, Cau Giay, Hanoi, Vietnam\\
\AdrHEPC \\
$^*$Email: dong.phungvan@phenikaa-uni.edu.vn} 

\begin{abstract}

The flipped trinification, a framework for unifying the 3-3-1 and left-right symmetries, has recently been proposed in order to solve profound questions, the weak parity violation and the number of families, besides the implication for neutrino mass generation and dark matter stability. In this work, we argue that this gauge-completion naturally provides flavor-changing neutral currents in both quark and lepton sectors. The quark flavor changing happens at the tree-level due to the nonuniversal couplings of $Z'_{L,R}$, while the lepton flavor changing $l\rightarrow l'\ga$ starts from the one loop level contributed significantly by the new charged currents of $Y_{L,R}$, which couple ordinary to exotic leptons. These effects disappear in the minimal left-right model, but are present in the framework characterizing a flipped trinification symmetry.

\end{abstract}

\pacs{12.60.-i}
\date{\today}

\maketitle

\section{\label{introduction} Introduction}

The experiments of neutrino oscillations caused by nonzero small neutrino masses and flavor mixing have provided the most important evidences that prove the new physics beyond the standard model~\cite{neutrino}. The compelling way to address the neutrino masses is to introduce right-handed neutrinos into the standard model, by which the neutrino mass generation is done by seesaw mechanisms \cite{seesaw}. The pioneering model that recognizes the seesaw mechanisms
is the minimal left-right symmetric model \cite{LR}, where the neutrino masses were predicted before the experimental confirmations. 

The minimal left-right symmetric model offers a possibility to understand the origin of the parity violation of weak interactions, but as the standard model it neither shows why there are only three fermion generations nor addresses dark matter stability that accounts for more than $25 \%$ mass-energy density of the universe \cite{darkrv}. Indeed, the lightest right-handed neutrino may have a keV mass responsible for warm dark matter, but it would overpopulate the universe due to gauge interactions, which require nonstandard dilution mechanisms \cite{lrdm1}. On the other hand, cold dark matter scenario that adds a new field as well as imposing a stabilizing symmetry remains to be arbitrary, ad hoc included \cite{lrdm2}.  
 
It is well established that the 3-3-1 model \cite{331} provides a potential solution to the generation number and addresses the issue of dark matter naturally \cite{3311}. Hence, we have recently proposed a theoretical model that unifies both the left-right and 3-3-1 symmetries, resulting in a $SU(3)_C\otimes SU(3)_L\otimes SU(3)_R \otimes U(1)_X$ gauge group, called flipped trinification \cite{3331DM} (for other interpretations, see \cite{3331}). This model inherits all nice features of both left-right and 3-3-1 models. Particularly, dark matter naturally exists which along with normal matter form gauge multiplets by the gauge symmetry, whereas the three generations emerge as a result of anomaly cancellation. Moreover, the origin of the matter parity and the dark matter stability are determined by a residual gauge symmetry. The new physics predicted occurs at TeV scale, giving rise to interesting signatures at current colliders.
 
An intriguing feature of the flipped trinification is that flavor violating interactions appear in both quark and lepton sectors. 
As a trinification symmetry is flipped, both left- and right-handed quark flavors transform differently under $SU(3)_{L,R}$. Consequently, they lead to tree-level flavor-changing neutral currents (FCNCs) that coupe to $Z'_{L,R}$, where the relevant observables after integrating out $Z'_{L,R}$ depend on both left- and right-handed quark mixing matrices.  Further, the discovery of neutrino oscillations suggests lepton flavor violation (LFV), but the charged LFV has never been observed. As the minimal left-right gauge symmetry is enlarged to trinification, the model predicts new non-Hermitian gauge bosons $Y_{L,R}$ that couple charged leptons to new heavy leptons. 
This is the main source for charged LFV processes $l\rightarrow l'\ga$ that are mediated by $Y_{L,R}$ in one-loop corrections, since the new leptons mix. Additionally, the contributions of $W_{L,R}$ due to the neutrino mixing and of new Higgs bosons to such charged LFV processes will be taken into account.  Moreover, the flipped trinification has scalar fields that couple both charged leptons and flavor change. This leads to tree-level charged LFV processes such as $\tau \rightarrow 3\mu (3e)$, $\mu \rightarrow 3e$, and so forth. 

Let us recall that due to the left-right symmetry, the model requires a bitriplet $\varphi$, two triplets $\chi_L, \chi_R$, and two sextets $\si_L, \si_R$ in order for the gauge symmetry breaking and mass generation. Since $\chi_L,\si_L$ have small vacuum expectation values (VEVs), 
their roles were ignored in the previous study \cite{3331DM}. In this work, we will turn on their effects when including the full scalar sector, which contribute to the gauge symmetry breaking pattern and mass spectra. We show that the VEVs of $\chi_L$ and $\si_L$ do not affect significantly the gauge boson masses, but the neutrino and Higgs spectra are modified. 

The rest of this paper is organized as follows. In Section \ref{model}, we reconsider the model with the complete scalar content. In Section \ref{FCNC}, we study the tree-level FCNCs and their contributions to neutral meson mixings, which are mediated by new gauge bosons $Z^\prime_L,Z^\prime_R$. In Section \ref{CLFV}, we present analytic expressions and numerical results for the specific charged LFV processes. Finally, we conclude this work in Sec.\ref{Con}.

\section{\label{model} A review of the model}

This section necessarily imposes $\chi_L, \si_L$ due to the left-right symmetry, which were omitted in the previous study for a mathematical simplicity \cite{3331DM}.  

\subsection{Symmetry and field content}

Left-right symmetrizing the 3-3-1 group \cite{331}, we obtain a gauge symmetry \be SU(3)_C\otimes SU(3)_L\otimes SU(3)_R \otimes U(1)_X,\ee which matches a flipped trinification and preserves the $SU(3)_L$ and $SU(3)_R$ interchange. The electric charge operator is given by  
\be
Q=T_{3L}+T_{3R}+\beta\left(T_{8L}+T_{8R}\right)+X,
\ee where $T_{iL,R}$ $(i=1,2,3,...,8)$ and $X$ are $SU(3)_{L,R}$ and $U(1)_X$ charges, respectively. The baryon minus lepton number is identified as   
\be 
\fr{1}{2}\left(B-L \right)=\beta\left(T_{8L}+T_{8R}\right)+X,
\ee
which is non-commutative, in contrast to the usual (Abelian) extensions. We further define a basic electric charge as $q=-(1+\sqrt{3}\beta)/2$.

Analogously, the fermion content is obtained from those of the 3-3-1 model by left-right symmetrization, which yields
  \be \psi_{aL} =
\left(\begin{array}{c}
               \nu_{aL}\\ e_{aL}\\ N^q_{aL}
\end{array}\right) \sim \left(1,3, 1,\frac{q-1}{3}\right),  \hs \hs  \psi_{aR} =
\left(\begin{array}{c}
               \nu_{aR}\\ e_{aR}\\ N^q_{aR}
\end{array}\right) \sim \left(1,1, 3,\frac{q-1}{3}\right),\ee
\be Q_{\al L}=\left(\begin{array}{c}
  d_{\al L}\\  -u_{\al L}\\  J^{-q-\frac{1}{3}}_{\al L}
\end{array}\right)\sim \left(3,3^*,1,-\frac{q}{3}\right), \hs \hs
Q_{\al R}=\left(\begin{array}{c}
  d_{\al R}\\  -u_{\al R}\\  J^{-q-\frac{1}{3}}_{\al R}
\end{array}\right)\sim \left(3,1,3^*,-\frac{q}{3}\right), \ee   \be  Q_{3L}= \left(\begin{array}{c} u_{3L}\\  d_{3L}\\ J^{q+\frac{2}{3}}_{3L} \end{array}\right)\sim
 \left(3,3,1,\frac{q+1}{3}\right), \hs \hs  Q_{3R}= \left(\begin{array}{c} u_{3R}\\  d_{3R}\\ J^{q+\frac{2}{3}}_{3R} \end{array}\right)\sim
 \left(3,1,3,\frac{q+1}{3}\right),\ee
where $a=1,2,3$ and $\al=1,2$ are generation indices. The model predicts new fermions $N_a, J_a$, besides the right-handed neutrinos $\nu_{aR}$. The fermion sector is more economical than that of the well-known trinification \cite{trinification}. In contrast to the trinification, the $SU(3)_L$ or $SU(3)_R$ anomaly cancellation requires the number of generations to match that of colors, and that the third quark generation transforms under $SU(3)_{L,R}$ differently from the first two quark generations, analogous to the 3-3-1 model \cite{331}.     

To break the gauge symmetry and generate the masses appropriately, the scalar multiplets are supplied as
\bea
\phi &=& \left(
\begin{array}{ccc}
 \phi_{ 11}^0 & \phi_{ 12}^+& \phi_{ 13}^{-q} \\
  \phi_{21}^- & \phi_{ 22}^{0} & \phi_{ 23}^{-1-q} \\
  \phi_{ 31}^{q}& \phi_{ 32}^{1+q}& \phi_{ 33}^{0} \\
\end{array}
\right)     \sim (1,3,3^*,0), \crn
 \chi_L &=& \left( \begin{array}{ccc}\chi_1^{-q}\\ \chi_2^{-q-1}\\ \chi_3^0 \end{array} \right)_L \sim\left(1,3,1,-\frac{2q+1}{3} \right), \crn
 \chi_R &=& \left( \begin{array}{ccc}\chi_1^{-q}\\ \chi_2^{-q-1}\\ \chi_3^0 \end{array} \right)_R \sim\left(1,1,3,-\frac{2q+1}{3} \right), \crn
 \sigma_L &=& \left(%
\begin{array}{ccc}
 \sigma_{11}^0 & \frac{\sigma_{12} ^-}{\sqrt{2}}& \frac{\sigma_{ 13}^{q}}{\sqrt{2}} \\
  \frac{\sigma_{12}^-}{\sqrt{2}} & \sigma_{22}^{--} &\frac{ \sigma_{23}^{q-1}}{\sqrt{2}} \\
 \frac{ \sigma_{13}^{q}}{\sqrt{2}}&\frac{ \sigma_{23}^{q-1}}{\sqrt{2}}& \sigma_{33}^{2q} \\
\end{array}%
\right)_L     \sim \left(1,6,1,\frac{2(q-1)}{3}\right), \crn
\sigma_R &=& \left(%
\begin{array}{ccc}
 \sigma_{11}^0 & \frac{\sigma_{12} ^-}{\sqrt{2}}& \frac{\sigma_{ 13}^{q}}{\sqrt{2}} \\
  \frac{\sigma_{12}^-}{\sqrt{2}} & \sigma_{22}^{--} &\frac{ \sigma_{23}^{q-1}}{\sqrt{2}} \\
 \frac{ \sigma_{13}^{q}}{\sqrt{2}}&\frac{ \sigma_{23}^{q-1}}{\sqrt{2}}& \sigma_{33}^{2q} \\
\end{array}%
\right)_R     \sim \left(1,1,6,\frac{2(q-1)}{3}\right),\eea
which reflect the left-right symmetry. The corresponding VEVs are given by 
\bea
\langle\phi \rangle&=&\frac{1}{\sqrt{2}} \left(%
\begin{array}{ccc}
u & 0 &0\\
 0 & u^\prime &0 \\
0& 0&w \\
\end{array}%
\right), \\  
\langle \chi_L \rangle &=&\frac{1}{\sqrt{2}} \left(%
\begin{array}{ccc}
	0\\
	0 \\
	w_L \\
\end{array}%
\right),\hs \langle \chi_R \rangle =\frac{1}{\sqrt{2}} \left(%
\begin{array}{ccc}
0\\
 0 \\
w_R \\
\end{array}%
\right), \label{vev2} \\ 
\langle \si_L \rangle &=&\frac{1}{\sqrt{2}} \left(%
\begin{array}{ccc}
	\Lambda_L & 0 &0\\
	0 & 0&0 \\
	0& 0&0 \\
\end{array}%
\right),\hs \langle \si_R \rangle =\frac{1}{\sqrt{2}} \left(%
\begin{array}{ccc}
\Lambda_R & 0 &0\\
 0 & 0&0 \\
0& 0&0 \\
\end{array}%
\right). \label{vev1}
  \eea   
  
As shown in \cite{3331DM}, the symmetry breaking is proceeded through several schemes, depending on the hierarchy arrangements of the VEVs. All the schemes lead to the existence of a residual discrete gauge symmetry that conserves every VEV, called matter parity \be W_P=(-1)^{3(B-L)+2s}=(-1)^{6[\beta(T_{8L}+T_{8R})+X]+2s}.\ee This parity ensures the stability of dark matter which is unified with normal matter in the gauge multiplets (see \cite{3331DM} for details of the dark sector and dark matter candidates). For consistency, we assume $\La_R, w_R, w \gg u,u^\prime \gg \La_L, w_L$, appropriate to the potential minimization. [Indeed, the minimization conditions imply $\La_L\simeq 0,\ w_L\simeq 0$, where the small nonzero values come from abnormal perturbative interactions, as seen in the next section]. This means that the flipped trinification is broken down to the standard model and matter parity, and then to the remnant $SU(3)_C\otimes U(1)_Q\otimes W_P$, where the left-right asymmetry is explicitly recoginzed at the electroweak phase due to $w\neq 0$, $w_R\neq w_L$ and $\La_R\neq \La_L$. 

\subsection{Fermion masses}
First, we consider the physical states and masses of fermions. They arise from  the Yukawa interactions as follows
\bea \mathcal{L}_{\mathrm{Yukawa}} & =& x_{ab}\left( \bar{\psi}^c_{aR} \sigma^\dagger_R \psi_{bR}+\bar{\psi}^c_{aL} \sigma^\dagger_L \psi_{bL}\right) + y_{ab} \bar{\psi}_{aL} \phi \psi_{b R}+\fr{z_{ab}}{M}\bar{\psi}_{aL}\chi_L \chi_R^* \psi_{bR} \crn
&&+ k_{33}\bar{Q}_{3L} \phi Q_{3R} +k_{\al \beta}\bar{Q}_{\al L} \phi^* Q_{\beta
R}+ \fr{k'_{33}}{M}\bar{Q}_{3L} \chi_L \chi_R^* Q_{3 R}+\fr{k'_{\al \beta}}{M} \bar{Q}_{\al L}\chi_L^* \chi_R Q_{\beta R}  \nonumber\\ 
&& + \fr{t_{3\al}}{M} \left(\bar{Q}_{3L}\phi \chi_R^* Q_{\al R}+\bar{Q}_{3R}\phi^* \chi_L^* Q_{\al L}\right) + \fr{t_{\al 3}}{M} \left(\bar{Q}_{\al L} \phi^* \chi_R Q_{3R}+\bar{Q}_{\al R}\phi \chi_{L}Q_{3L} \right)\crn
&& +H.c.,  \label{yukawa}\eea where $M$ is a new physics scale that defines the effective interactions. The left-right symmetry demands that the couplings $y,z,k,k'$ are Hermitian, whereas $x,t$ are generic. 

After the symmetry breaking, the Yukawa Lagrangian yields  fermion masses. The new leptons 
get a large mass at the new physical scale as follows
\bea \mathcal{L}_{\mathrm{mass}}^N = \left(\fr{y_{ab}}{\sqrt{2}}w+\fr{z_{ab}}{2M}w_L w_R\right) \bar{N}_{aL} N_{bR} +H.c.
\label{maN}\eea
The ordinary charged leptons obtain a mass at the weak scale,
\bea
\mathcal{L}_{\mathrm{mass}}^l= \left(\fr{y_{ab}}{\sqrt{2}}u^\prime \right) \bar{l}_{aL} l_{bR}+ H.c.
\label{mal}\eea
Note that the new leptons do not mix with the ordinary leptons due to the matter parity conservation. If 
neglecting the effective interactions, they have the same mixing matrices. 

The Lagrangian (\ref{yukawa}) allows neutrinos having both kinds of mass terms: Dirac and Majorana. In the basis $(\nu_L, \nu^c_R)$, the neutrino mass matrix is given by 
\bea
\mathcal{M}_\nu =\left(%
\begin{array}{cc}
	M_\nu^L & M_\nu^D \\
	(M_\nu^D)^T & M_\nu^R \\
\end{array}%
\right),
\eea
where the explicit forms of $M_\nu^D, M_\nu^L, M_\nu^R$ are
\bea
\left(M_\nu^D \right)_{ab} =-\fr{y_{ab}}{\sqrt{2}} u, \hs \left(M_\nu^L \right)_{ab}=-\sqrt{2}x_{ab}\La_L, \hs
\left(M_\nu^R \right)_{ab}=-\sqrt{2}x_{ab}\La_R.
\eea Because of the condition $\La_L\ll u\ll \La_R$, the active neutrinos ($\sim \nu_L$) gain small masses via the seesaw mechanisms, 
\be M_\nu\simeq -\sqrt{2} x \La_L+\fr{1}{2\sqrt{2}}yx^{-1}y^T\fr{u^2}{\La_R},\ee whereas the sterile neutrinos ($\sim \nu_R$) have large masses at $\La_R$ scale, $M'_\nu\simeq -\sqrt{2}x\La_R$.   

The exotic quarks do not mix with ordinary quarks due to the matter parity conservation and have the mass terms given from (\ref{yukawa}) by
\bea
\mathcal{L}_{\mathrm{mass}}^J=\left(\fr{k_{33}}{\sqrt{2}}w+\fr{k^\prime_{33}}{2M}w_L w_R \right)\bar{J}_{3L} J_{3R}+\left(\fr{k_{\alpha \beta}}{\sqrt{2}}w+\fr{k^\prime_{\alpha \beta}}{2M}w_L w_R \right)\bar{J}_{\al L} J_{\beta R}+H.c.,  
\eea which are all at the new physics scale.

Denoting $u_{L,R}=\left(u_1,u_2,u_2\right)^T_{L,R}$ and $d_{L,R}=\left(d_1,d_2,d_3 \right)^T_{L,R}$, the ordinary quarks achieve mass terms
\bea
\mathcal{L}^{u,d}=-\bar{u}_L\mathcal{M}^u u_R-\bar{d}_L\mathcal{M}^d d_R+H.c.,
\eea
where
\bea
\mathcal{M}^u&=&-\fr{1}{\sqrt{2}} \left(\begin{array}{ccc} k_{11}u^\prime& k_{12}u^\prime &\fr{-t_{13}}{M\sqrt{2}}u^\prime\left(w_L+w_R \right)\\ k_{21}u^\prime& k_{22}u^\prime &\fr{-t_{23}}{M\sqrt{2}}u^\prime\left(w_L+w_R \right) \\ \fr{t_{31}}{M\sqrt{2}}u\left(w_L+w_R \right)&\fr{t_{32}}{M\sqrt{2}}u\left(w_L+w_R \right)&k_{33}u\end{array} \right), \crn \nonumber  \\ 
\mathcal{M}^d&=&-\fr{1}{\sqrt{2}} \left(\begin{array}{ccc} k_{11}u& k_{12}u &\fr{-t_{13}}{M\sqrt{2}}u\left(w_L+w_R \right)\\ k_{21}u& k_{22}u &\fr{-t_{23}}{M\sqrt{2}}u\left(w_L+w_R \right) \\ \fr{t_{31}}{M\sqrt{2}}u^\prime\left(w_L+w_R \right)&\fr{t_{32}}{M\sqrt{2}}u^\prime\left(w_L+w_R \right)&k_{33}u^\prime\end{array} \right).
\eea
Applying bi-unitary transformations, the mass matrices can be diagonalized 
as 
\bea
M^d=V_{dL}^\dag \mathcal{M}^d V_{dR}, \hs M^u=V_{uL}^\dag \mathcal{M}^u V_{uR},
\eea 
where $M^{u}, M^d$ are diagonal matrices that consist of respective physical quark masses at the weak scale, given that $M\sim w_{R}$. Note that the mass eigenstates $u'=(u,c,t)^T$ and $d'=(d,s,b)^T$ are related to the gauge states by $u_{L,R}=V_{uL,R}u^\prime_{L,R}$ and $d_{L,R}=V_{dL,R}d^\prime_{L,R}$.

\subsection{Gauge boson masses \label{gaugemass}}
The presence of the scalar multiplets $\si_L, \chi_L$
does not significantly change the mass spectrum of the gauge bosons that was derived in \cite{3331DM}. Hereafter, we summarize the main results of the gauge sector. The gauge bosons $W_L,W_R$ slightly mix, which yield eigenstates \be W_{1 }= c_\xi W_{L } -s_\xi W_{R },\hs W_{2 }= s_\xi W_{L } +c_\xi W_{R },\ee where the mixing angle $\xi$ is defined by \be t_{2 \xi} = \frac{4 t_R u u'}{2  \Lambda _L^2-2 \Lambda _R^2 t_R^2-\left(t_R^2-1\right) (u^2+ u'^2)}\simeq -\fr{2uu'}{t_R\La^2_R}\ll 1.\ee The $W_{1,2}$ masses are given by
\bea
 m_{W_1}^2 && \simeq \frac{g_L^2}{4}\left[u^2+u^{\prime 2}+2\La_L^2-\fr{4t_R^2 u^2 u^{\prime 2}}{(t_R^2-1) (u^2+u^{\prime 2}) +2t_R^2 \Lambda^2_R-2\La_L^2} \right]\simeq \fr{g^2_L}{4}(u^2+u'^2), \\ m_{W_2}^2  && \simeq \frac{g_R^2}{4} \left[u^2+u^{\prime 2}+2\La_R^2+\fr{4 t^2_R u^2 u^{\prime 2}}{(t_R^2-1) (u^2+u^{\prime 2}) +2t_R^2 \Lambda^2_R-2\La_L^2} \right]\simeq \fr{g^2_R}{2}\La^2_R, \eea where $g_L,g_R$ are $SU(3)_{L,R}$ couplings respectively, which match $t_R\equiv g_R/g_L=1$ at the flipped trinification scale due to the left-right symmetry. At the low energy, they may separate, $t_R\neq 1$, due to the different contributions to the running couplings. $W_1$ is identical to the standard model $W$ boson, implying $u^2+u'^2=(246\ \mathrm{GeV})^2$, while $W_2$ is new. 
 
Besides, the model predicts new non-Hermitian gauge bosons $X_{L,R}^{\pm q}$ and $Y^{\pm (q+1)}_{L,R}$ that couple to the charges $T_4\mp i T_5$ and $T_6\mp i T_7$, respectively. The physical states are \bea && X^{\pm q}_1 = c_{\xi_1} X_L^{\pm q} -s_{\xi_1} X_R^{\pm q},\hs X^{\pm q}_2 = s_{\xi_1} X_L^{\pm q} +c _{\xi_1} X_R^{\pm q},\\ && Y_{1}^{\pm (1+q)}= c_{\xi_2}Y_L^{\pm(1+q)}-s_{\xi_2}Y_R^{\pm(1+q)},\hs Y_{2}^{\pm (1+q)}= s_{\xi_2}Y_L^{\pm(1+q)}+c_{\xi_2}Y_R^{\pm(1+q)}.\eea Here,
the mixing angles $\xi_1, \xi_2$ are obtained as \bea 
&& t_{2 \xi_1}=\frac{4 t_R u w}{u^2 + w^2 + w_L^2 +2 \Lambda_L^2- t_R^2(u^2+w^2 + w_R^2  + 2 \Lambda_R^2)}\sim \fr{u}{w},\\
&& t_{2 \xi_2} = \fr{4 t_R u' w}{u'^{2}+w^2 +w_L^2-
	t_R^2(u'^2+w_R^{ 2}+w^2)}\sim \fr{u'}{w}.\eea
And, the gauge boson masses are given by 
\bea
m_{X_1}^2 &=& \frac{g_L^2}{4} \left[u^2+w^2+w_L^2+2\Lambda_L^2 + \frac{ 4t_R^2u^2w^2}{u^2+w^2+w_L^2+2\La_L^2-t_R^2(u^2+w^2+w^{2}_R+2 \Lambda^2_R) }\right]\crn
&\simeq&\fr{g^2_L}{4}w^2, \\
m_{X_2}^2  &=& \frac{g_R^2}{4} \left[u^2+w^2+w_R^2+2 \Lambda_R^2 - \frac{ 4u^2w^2}{u^2+w^2+w_L^2+2\La_L^2-t_R^2(u^2+w^2+w^{2}_R+2 \Lambda^2_R) }\right]\crn
&\simeq& \fr{g^2_R}{4}(w^2+w^2_R+2\La^2_R), \\
m^2_{Y_1} &=& \fr{g_L^2}{4} \left[u^{\prime 2}+w^2 +w_L^2+\fr{4 t_R^2 u^{\prime 2}w^2}{u^{\prime 2}+w^2+w_L^2 -t_R^2(u^{\prime 2}+w^2+w_R^{ 2})}\right]\simeq\fr{g^2_L}{4}w^2, \\
m^2_{Y_2} &=& \fr{g_R^2}{4} \left[u^{\prime 2}+w^2+w_R^{2} -\fr{4  u^{\prime 2}w^2}{u^{\prime 2}+w^2+w_L^2 -t_R^2(u^{\prime 2}+w^2+w_R^{ 2})}\right]\simeq \fr{g^2_R}{4}(w^2+w^2_R).
\eea

The neutral gauge bosons $A_{3L,R},A_{8L,R},B$, that couple to the charges $T_{3L,R}, T_{8L,R}, X$ respectively, mix via a $5 \times 5$ mass matrix, given 
in Appendix \ref{Appendix}. The photon field is \be A = s_W A_{3L}+c_W \left(\fr{t_W}{t_R}A_{3R}+\beta t_W A_{8L}+\beta \fr{t_W}{t_R}A_{8R}+\fr{t_W}{t_X}B\right),\ee 
which is massless, where $t_X\equiv g_X/g_L$ is $U(1)_X/SU(3)_L$ coupling ratio. The sine of the Weinberg angle is $ s_W = t_X t_R/\sqrt{t_X^2(1+\beta^2)+t_R^2(1+t_X^2(1+\beta^2))}$, obtained by matching the electromagnetic gauge coupling \cite{donglong2005}. As usual, the standard model $Z$ boson is given orthogonally to $A$ by \be Z_L = c_W A_{3L}-s_W \left(\fr{t_W}{t_R}A_{3R}+\beta t_W A_{8L}+\beta \fr{t_W}{t_R}A_{8R}+\fr{t_W}{t_X}B\right).\ee New neutral gauge bosons take the forms that are orthogonal to both $A$ and $Z_L$, i.e. to the $U(1)_Y$ gauge field in the parentheses, \bea && Z^\prime_L=\varsigma_1  t_X t_W \beta A_{3R}-\fr{t_W}{\varsigma_1 t_X t_R}A_{8L}+\varsigma_1  t_X t_W \beta^2 A_{8R}+\varsigma_1 t_R  t_W \beta B,\\  
&&Z_R=-\fr{\varsigma_1}{\varsigma}A_{3R}+\varsigma \varsigma_1  t_X^2 \beta A_{8R}+\varsigma \varsigma_1 t_X t_R B,\\  
&&Z^\prime_R=\varsigma (t_R A_{8R} -t_X \beta B),\eea where $ \varsigma =1/\sqrt{t_R^2+\beta^2 t_X^2}$ and $\varsigma_1=1/\sqrt{t_R^2+(1+\beta^2)t_X^2}$.

In the new basis ($A, Z_L, Z'_L, Z_R, Z'_R$), $A$ is decoupled, while $Z_L$ infinitesimally mixes with $(Z_L^\prime, Z_R, Z_R^\prime)$ where the relevant mixing angles are suppressed by $(u,u')^2/(w,w_{R},\La_R)^2\ll 1$. Neglecting the mixing, $Z_L$ is a physical field and decoupled as the photon. We are left with diagonalizing the mass matrix of $(Z_L^\prime, Z_R, Z_R^\prime)$, which yields the eigenstates $\mathcal{Z}_L^\prime, \mathcal{Z}_R, \mathcal{Z}_R^\prime$ and corresponding masses as 
\bea
\mathcal{Z}^\prime_L &\simeq& Z_L^\prime, \hs \mathcal{Z}_R \simeq c_{\xi_3} Z_R -s_{\xi_3} Z_R^\prime, \hs  \mathcal{Z}_R^\prime \simeq s_{\xi_3} Z_R +c_{\xi_3} Z_R^\prime,\\
m^2_{\mathcal{Z}_L^\prime} &\simeq& \frac{g_L^2}{3}\frac{(1+\varsigma_1^2t_R^2t_X^2\beta^2)^2t_W^2 w^2}{\varsigma_1^2 t_R^2 t_X^2}, \\
m^2_{\mathcal{Z}_R} &\simeq&\fr{g_L^2}{3} \frac{3w^{2}_R[t_R^2+t_X^2(1+\beta^2)]^2+w^2\left[\sqrt{3}t_R^2+\left(\sqrt{3}+\beta\right)t_X^2\right]^2}{\varsigma_1^{-2}[4+(\sqrt{3} +\beta)^2(t_X^2/t_R^2)]}, \\
m^2_{\mathcal{Z_R^\prime}} & \simeq & \frac{g_L^2}{3}\left[ 4t_R^2+t_X^2(\sqrt{3}+ \beta)^2\right] \Lambda^2_R,
\eea provided that $\La_R\gg w,w_{R}$, where
the $Z_R$-$Z'_R$ mixing angle is finite,
\bea
t_{2\xi_3} &=& \fr{2t_R\left [\sqrt{3}t_R^2+ \beta(3+\sqrt{3}\beta)t_X^2 \right ]\sqrt{t_R^2+t_X^2(1+\beta^2)}}{2t_R^4+t_R^2t_X^2(\sqrt{3}-\beta )^2-\beta^2(\sqrt{3}+\beta )^2t_X^4}.
\eea Analogously, we can diagonalize the mass matrix for the case $\La_R\ll w,w_{R}$, where $Z_R$ is decoupled, while $Z'_{L,R}$ finitely mix. For the case $\La_R\sim w,w_{R}$, all the gauge bosons $Z'_{L,R},Z_R$ finitely mix, which can be parameterized by the Euler angles. Note that $w,w_R$ are always taken in the same order, since they simultaneously break $SU(3)_L\otimes SU(3)_R\rightarrow SU(2)_L\otimes SU(2)_R$ and correspondingly reduce the left-right symmetry.  
 
\subsection{Higgs masses}
Let us rewrite the scalar potential that includes the full scalar content. The full scalar potential takes the form, 
$V = V_\phi+V_\chi+V_\sigma+V_{\mathrm{mix}}$, where
 \bea V_\phi &=& \mu^2_\phi \Tr(\phi^\dagger \phi) +\la_1 [\Tr(\phi^\dagger \phi)]^2+\la_2 \Tr[(\phi^\dagger \phi)^2],\nonumber \\
 V_\chi &=& \mu^2_{\chi}\left[(\chi^\dagger_L \chi_L+ \chi^\dagger_R \chi_R\right] + \ka_{1} \left[(\chi_L^\dagger \chi_L)^2+ (\chi_R^\dagger \chi_R)^2 \right]+\ka_2 (\chi_L^\dagger\chi_L)(\chi_R^\dagger \chi_R), \nonumber \\
 V_\sigma &=& \mu^2_{\sigma}\left[ \Tr\sigma_L^\dagger \sigma_L+ \Tr\sigma_R^\dagger \sigma_R \right]+\rho_1 \left\{ [\Tr(\sigma^\dagger_L \sigma_L)]^2 +[\Tr(\sigma^\dagger_R \sigma_R)]^2\right\}\nonumber \\ &+&\rho_2 \left\{\Tr[(\sigma^\dagger_L \sigma_L)^2]+\Tr[(\sigma^\dagger_R \sigma_R)^2] \right\}+\rho_3\Tr[\sigma^\dagger_L \sigma_L]\Tr[\sigma^\dagger_R \sigma_R], \nonumber \\
 V_{\mathrm{mix}} &=& \zeta_1 \left[\chi^\dagger_L \chi_L+\chi^\dagger_R \chi_R\right]\Tr(\phi^\dagger \phi)\nonumber+\zeta_2 \left[ \chi^\dagger_L \chi_L\Tr(\sigma^\dagger_L \sigma_L)+\chi^\dagger_R \chi_R\Tr(\sigma^\dagger_R \sigma_R) \right] \nonumber \\ &+&\zeta_3 \left[\chi_L^\dagger \sigma_L \sigma_L^\dagger \chi_L+ \chi_R^\dagger \sigma_R \sigma_R^\dagger \chi_R\right]+
 \zeta_4\left[ \chi_L^\dagger \chi_L \Tr(\sigma^\dagger_R \sigma_R)+\chi_R^\dagger \chi_R \Tr(\sigma^\dagger_L \sigma_L)\right]\nonumber \\ &+& \zeta_5 \left[ \chi_L^\dagger \phi\phi^\dagger\chi_L+\chi_R^\dagger \phi^\dagger\phi\chi_R\right] +\zeta_6\left[\Tr(\sigma^\dagger_L \sigma_L) +\Tr(\sigma^\dagger_R \sigma_R) \right] \Tr(\phi^\dagger \phi) \nonumber \\ &+&\zeta_7\left[\Tr(\phi^\dagger \phi \sigma_R \sigma^\dagger_R)+\Tr(\phi \phi^\dagger \sigma_L \sigma^\dagger_L) \right] +\zeta_8 \left[\epsilon^{ijk} \epsilon_{\al \beta \gamma}\chi_{Li} \phi_j^\al \phi_k^\beta \chi_R^{\dagger\gamma}+H.c. \right]\nonumber \\
 &+& \zeta_9 [\sigma_{Lij}\phi^{\dagger i}_\al \phi^{\dagger j}_\beta \sigma^{\dagger\al\beta}_R+H.c.]+\left[f_1 \epsilon^{ijk} \epsilon_{\al \beta \gamma}\phi_i ^\al \phi_j^\beta \phi_k^\gamma+f_2\chi_L^\dagger \phi \chi_R+ H.c.\right].
  \eea Here the interactions $\zeta_{8,9}$, $f_{1,2}$ are abnormal and subdominant since they can be removed by a global symmetry $U(1)$ that nontrivially transforms any one of the fields. 
  
Expanding the neutral scalar fields around their VEVs, we  find minimization conditions, mass terms, and interactions. The mass terms are divided as $V_{\mathrm{mass}}=V_S+V_A+V_{\mathrm{charged}}$, where $V_S,V_A$ include those of CP even and CP odd scalar fields, respectively, whereas $V_{\mathrm{charged}}$ consists of those of the charged scalars. Considering $q+1$-charged scalars, four fields $(\phi_{23}^{\pm (q+1)}, \phi_{32}^{\pm (q+1)},  \chi_{R}^{\pm (q+1)}, \chi_{L}^{\pm (q+1)})$ mix via a $4 \times 4$ matrix, which by diagonalization provides two massless Goldstone bosons, $G_{Y_L}^{\pm(q+1)}, G_{Y_R}^{\pm(q+1)}$ and two massive Higgs fields, $\mathcal{H}_{1}^{\pm(q+1)}, \mathcal{H}_{2}^{\pm(q+1)}$. These states are related to the gauge states by  \be (\phi_{23}^{\pm (q+1)}\ \phi_{32}^{\pm (q+1)}\ \chi_{R}^{\pm (q+1)}\ \chi_{L}^{\pm (q+1)})^T=\mathcal{P}(G_{Y_L}^{\pm (q+1)}\ G_{Y_R}^{\pm (q+1)} \mathcal{H}_1^{\pm (q+1)}\ \mathcal{H}_2^{ \pm (q+1)})^T,\ee
where
\bea
\mathcal{P}&\simeq&\left(
\begin{array}{cccc}
	0 &  -\fr{\sqrt{(u^{\prime 2}-w^2)^2+w^2w_R^2}}{{\sqrt{(u^{\prime 2}-w^2)^2+(w^2+u^{\prime 2})w_R^2}}} &0 & \frac{u^{\prime 2}w_{R}}{\sqrt{(u^{\prime 2}-w^2)^2+(w^2+u^{\prime 2})w_R^2}}  \\
	\fr{w^2-u^{\prime 2}}{\sqrt{u^{\prime 2}-w^2)^2+w^2w_R^2}} & \frac{u^{\prime }ww_{R}}{\sqrt{((u^{\prime 2}-w^2)^2+w^2w_R^2))((u^{\prime 2}-w^2)^2+(w^2+u^{\prime 2})w_R^2)}} & 0 & \frac{w w_{R}}{\sqrt{(u^{\prime 2}-w^2)^2+(w^2+wu^{\prime 2})w_R^2}}    \\
	\fr{w w_R c_{\xi_4}}{\sqrt{u^{\prime 2}-w^2)^2+w^2w_R^2}} & \frac{u^{\prime} w_R(u^{\prime 2}-w^2)c_{\xi_4}}{\sqrt{((u^{\prime 2}-w^2)^2+w^2w_R^2))((u^{\prime 2}-w^2)^2+(w^2+u^{\prime 2})w_R^2)}}& -s_{\xi_4} & -\frac{(w^2-u^{\prime 2})c_{\xi_4}}{\sqrt{(u^{\prime 2}-w^2)^2+(w^2+wu^{\prime 2})w_R^2}} \\
	-\fr{w w_R s_{\xi_4}}{\sqrt{u^{\prime 2}-w^2)^2+w^2w_R^2}} & -\frac{u^{\prime} w_R(u^{\prime 2}-w^2)s_{\xi_4}}{\sqrt{((u^{\prime 2}-w^2)^2+w^2w_R^2))((u^{\prime 2}-w^2)^2+(w^2+u^{\prime 2})w_R^2)}} & -c_{\xi_4}&  \frac{(w^2-u^{\prime 2})s_{\xi_4}}{\sqrt{(u^{\prime 2}-w^2)^2+(w^2+wu^{\prime 2})w_R^2}}
\end{array}
\right),\nonumber
\eea
where the mixing angle $\xi_4$ is defined by
\bea
\tan 2\xi_4=t_{2\xi_4} & \simeq & -\frac{2 \zeta_9 u \sqrt{u^4+u^{\prime 4}+2u^{\prime 2}\Lambda_R^2-2u^2(u^{\prime 2}-\Lambda_R^2)}}{\zeta_7 \Lambda_R^2 u^{\prime}}.
\eea

Concerning the singly-charged scalars, the model contains two massless Goldstone bosons $(G_{W_L}^\pm, G_{W_R}^\pm)$, which are eaten by $W_{L}^\pm,W_R^\pm$ respectively, and two physical massive fields $H_{1}^\pm, H_{2}^\pm$. 
They are related to the gauge states 
through $(\sigma_{12R}^{\pm}, \phi_{12}^{\pm}, \phi_{21}^{\pm}, \sigma_{12L}^{\pm})^T=\mathcal{K}(G_{W_L}^{\pm},  G_{W_R}^{\pm},  H_1^{\pm},  H_2^{\pm})^T$,
where
\bea
\mathcal{K}&\simeq&\left(
\begin{array}{cccc}
	0 &  -\fr{\sqrt{2(u^2+u^{\prime 2})}\Lambda_{R}}{\sqrt{(u^{\prime 2}-u^2)^2+2\Lambda_R^2(u^2+u^{\prime 2})}} &-\frac{(u^{\prime 2}-u^2)s_{\xi_5}}{\sqrt{(u^{\prime 2}-u^2)^2+2\Lambda_R^2(u^2+u^{\prime 2})}} & \frac{(u^{\prime 2}-u^2)c_{\xi_5}}{\sqrt{(u^{\prime 2}-u^2)^2+2\Lambda_R^2(u^2+u^{\prime 2})}}  \\
	\fr{u^{\prime}}{\sqrt{u^2+u^{\prime 2}}} & -\fr{u(u^2-u^{\prime 2})}{\sqrt{u^2+u^{\prime 2}}\sqrt{(u^{\prime 2}-u^2)^2+2\Lambda_R^2(u^2+u^{\prime 2})}}& -\fr{\sqrt{2}u\Lambda_{R}s_{\xi_5}}{\sqrt{(u^{\prime 2}-u^2)^2+2\Lambda_R^2(u^2+u^{\prime 2})}} & \fr{\sqrt{2}u\Lambda_{R}c_{\xi_5}}{\sqrt{(u^{\prime 2}-u^2)^2+2\Lambda_R^2(u^2+u^{\prime 2})}}   \\
	-\fr{u}{\sqrt{u^2+u^{\prime 2}}} & \fr{u^{\prime}(u^{\prime 2}-u^2)}{\sqrt{u^2+u^{\prime 2}}\sqrt{(u^{\prime 2}-u^2)^2+2\Lambda_R^2(u^2+u^{\prime 2})}}& -\fr{\sqrt{2}u^{\prime}\Lambda_{R}s_{\xi_5}}{\sqrt{(u^{\prime 2}-u^2)^2+2\Lambda_R^2(u^2+u^{\prime 2})}} & \fr{\sqrt{2}u^{\prime}\Lambda_{R}c_{\xi_5}}{\sqrt{(u^{\prime 2}-u^2)^2+2\Lambda_R^2(u^2+u^{\prime 2})}} \\
	0 & 0 & c_{\xi_5}& s_{\xi_5}
\end{array}
\right),\nonumber
\eea
with the mixing angel $\xi_5$ defined by
\bea
t_{2\xi_5} & \simeq & -\fr{{2u^{\prime}ww_Lw_R\sqrt{(w^2+w_L^2)(w^2+w_R^2)}}(w_L^2+w_R^2+2(w^2-u^{\prime 2}))}{(w_L^2-w_R^2)(u^{\prime 4}w^2+u^{\prime 2} w_L^2 w_R^2-w^2(w^2+w_L^2)(w^2+w_R^2))}.
\eea

The model also contains two heavy doubly-charged scalars $H_1^{\pm \pm}$ and $H_2^{\pm \pm}$, defined by
\bea
H^{\pm \pm}_{1}= c_{\xi_7}\sigma_{22R}^{\pm \pm}-s_{\xi_7}\sigma_{22L}^{\pm \pm}, \hs \hs H^{\pm \pm}_{2}&=& s_{\xi_7}\sigma_{22R}^{\pm \pm}+c_{\xi_7}\sigma_{22L}^{\pm \pm},
\eea
where $t_{2\xi_7}=\fr{2u^{\prime 2}\zeta_9 \Lambda_L \Lambda_R}{(\Lambda_L^2-\Lambda_R^2)(-u^2\zeta_9+2\Lambda_L\Lambda_R)}$.
Due to the limit $\La_R \gg \La_L$, the mixing angle $\xi_7 \simeq 0$, hence $\sigma_{22L},\sigma_{22R}$ are physical states by themselves. 

For the neutral scalars, they split into two parts: CP-odd and CP-even. The model contains only a light CP-even neutral scalar that is identified as the standard model Higgs boson, while the other CP-even states achieve large masses at the new physical scale. Additionally, the CP-odd part contains
four massless Goldstone bosons, which are correspondingly eaten by the eleven massive gauge
bosons $Z,Z_R, Z_L^\prime, Z_R^\prime$, and three heavy scalar states.

\section{\label{FCNC} FCNC}

As mentioned, the tree-level FCNCs arise due to the discrimination of quark generations, i.e. 
the third generations of left- and right-handed quarks $Q_{3L,R}$ transform differently from the first two $Q_{\al L,R}$ under $SU(3)_{L,R}\otimes U(1)_X$ gauge symmetry, respectively. Hence, the neutral currents will change ordinary quark flavors that nonuniversally couple to $T_{8L,R}$, since $X$ is related to $T_{8L,R}$ by the electric charge operator and that $Q,T_{3L,R}$ conserve every flavor. 

Indeed, with the aid of $X=Q-\left(T_{3L}+T_{3R} \right)-\beta \left( T_{8L}+T_{8R}\right)$, the neutral currents of quarks take the form
\bea 
\mathcal{L}_{NC}=-\bar{Q}_{L,R}\ga_\mu\left[g_{L,R}\left (T_{3L,R}A_{3L,R}^\mu+T_{8L,R}A_{8L,R}^\mu \right) +g_X\left(Q-T_{3L,R}-\beta T_{8L,R} \right)B^\mu\right]Q_{L,R},
\eea
where $Q_{L,R}$ are summed over all the quark multiplets. All the terms coupled to $T_{3L,R}, Q$ do not flavor change, 
because $u_{L,R}, d_{L,R}$ are identical under such charges. Hence, the FCNCs exist only for the terms that couple to $T_{8L,R}$,
\bea
\mathcal{L}_{NC}\supset-\sum_{a=1}^3 \bar{Q}_{aL,R}\ga_\mu T_{8L,R} Q_{aL, R}\left(g_{L,R}A_{8L,R}^\mu-\beta g_X B^\mu \right). \label{FCNC2}
\eea   

In the basis $(Z_{L}^\prime, Z_R, Z_R^\prime)$, the Lagrangian (\ref{FCNC2}) is rewritten as 
\bea
\mathcal{L}_{NC}\supset-\bar{q}^\prime_L\ga_\mu \left(V^\dag_{qL}T_{8q}V_{qL}  \right)q^\prime_L\left[g_1 Z^{\prime\mu}_L+g_2 Z_R^\mu+g_3 Z_{R}^{\prime \mu}   \right]  -\bar{q}^\prime_R\ga_\mu \left(V^\dag_{qL}T_{8q}V_{qR}\right)q^\prime_R  g_4 Z^{\prime \mu}_R, 
\label{FCNC3a}\eea
where we denote $q^\prime$ either $u^\prime$ or $d^\prime$, $T_{8q}=\fr{1}{2\sqrt{3}}\mathrm{diag}\left(1,1,-1\right)$, $g_1=g_Lt_Rt_X \cot \theta_W\varsigma_1$, $g_2=g_L\beta t_Rt_X^2  \varsigma \varsigma_1$, $g_3=-g_Lt_X^2\beta^2 \varsigma$, and $g_4=-\fr{g_L}{\varsigma}$. 

Taking, for instance, the limit $\La_R >w,w_R$ and changing to the mass basis, we obtain
\bea
\mathcal{L}_{NC}\supset&&-\bar{q}^\prime_L\ga_\mu \left(V^\dag_{qL}T_{8q}V_{qL}  \right)q^\prime_L\left[g_1 Z^{\prime\mu}_L+\left(g_2c_{\xi_3}-g_3s_{\xi_3} \right)\mathcal{Z}^{\mu}_R  +\left(g_2 s_{\xi_3}+g_3 c_{\xi_3}\right)\mathcal{Z}^{\prime\mu}_R \right] \nonumber \\  && -\bar{q}^\prime_R\ga_\mu \left(V^\dag_{qL}T_{8q}V_{qR}\right)q^\prime_R  g_4 \left(-s_{\xi_3}\mathcal{Z}_R^\mu+c_{\xi_3}\mathcal{Z}_{R}^{\prime \mu} \right). 
\label{FCNC3}\eea

It is noted that since $g_R A_{8R}-\beta g_X B \sim Z'_R$ and the large mixing $Z'_R$-$Z_R$, both $\mathcal{Z}'_R$ and $\mathcal{Z}_R$ contribute, whereas $g_L A_{8L}-\beta g_X B$ composes $Z'_L$ and these fields as it is not orthogonal to $Z'_R$. Consequently, the three fields $Z'_L, \mathcal{Z}_R, \mathcal{Z}_R^\prime$ dominantly couple to the tree-level FCNCs,  
\bea
\mathcal{L}_{FCNC}&=&\fr{1}{\sqrt{3}}\bar{q^\prime}_{iL}\ga_\mu q^\prime_{jL}\left( V^*_{qL}\right)_{3i}\left(V_{qL}\right)_{3j}\left[g_1 Z^{\prime\mu}_L+\left(g_2c_{\xi_3}-g_3s_{\xi_3} \right)\mathcal{Z}^{\mu}_R  +\left(g_2 s_{\xi_3}+g_3 c_{\xi_3}\right)\mathcal{Z}^{\prime\mu}_R \right]\nonumber \\
&&+\fr{1}{\sqrt{3}}\bar{q^\prime}_{iL}\ga_\mu q^\prime_{jL}\left( V^*_{qR}\right)_{3i}\left(V_{qR}\right)_{3j}g_4 \left(-s_{\xi_3}\mathcal{Z}_R^\mu+c_{\xi_3}\mathcal{Z}_{R}^{\prime \mu} \right),
\eea
for $i \neq j$. The new observation is that $Z_R$ flavor changes due to the large mixing with $Z'_R$, in contrast to the minimal left-right symmetric model. 

Integrating the heavy gauge bosons $\mathcal{Z}_R, \mathcal{Z}'_R,Z_L^\prime$ out, we determine the effective Lagrangian that describes the meson mixings,
\bea
\mathcal{L}^{eff}_{FCNC}&&=-\Upsilon^{ij}_L \left(\bar{q^\prime}_{iL} \ga_\mu q^\prime_{jL} \right)^2-\Upsilon^{ij}_R \left(\bar{q^\prime}_{iR} \ga_\mu q^\prime_{jR} \right)^2,
\eea
where 
\bea
\Upsilon^{ij}_L&&=\frac{1}{3}\left[\left( V^*_{qL}\right)^{3i}\left(V_{qL}\right)^{3j} \right]^2\left[\fr{g_1^2}{m_{Z_L^\prime}^2}+\fr{\left(g_2c_{\xi_3}-g_3s_{\xi_3} \right)^2}{m_{\mathcal{Z}_R}^2}
+\fr{\left(g_2 s_{\xi_3}+g_3 c_{\xi_3}\right)^2}{m_{\mathcal{Z}^\prime_R}^2}\right], \\  \Upsilon^{ij}_R&&=\frac{1}{3}\left[\left( V^*_{qR}\right)^{3i}\left(V_{qR}\right)^{3j} \right]^2\left[\fr{g_4^2s_{\xi_3}^2}{m_{\mathcal{Z}_R}^2}
+\fr{g_4^2c_{\xi_3}^2}{m_{\mathcal{Z}^\prime_R}^2}\right].
\label{FCNC4}\eea

Generally, the fields $Z_L,Z_R,Z_R^\prime$ mix via a ${3\times 3}$ mass matrix, as given in Appendix \ref{Appendix}. In this case, the mass eigenstates, $\mathcal{V}\equiv (\mathcal{Z}_1,\mathcal{Z}_2,\mathcal{Z}_3)$, are related to  $V\equiv (Z'_L,Z_R,Z_R^{\prime})$ by $V =U^Z\mathcal{V} $. Therefore, the couplings given in Eq. (\ref{FCNC4}) are generalized by
\bea
\Upsilon^{\prime ij}_L&=&\frac{1}{3}\left[\left( V^*_{qL}\right)^{3i}\left(V_{qL}\right)^{3j} \right]^2\left[ \fr{(g_1 U^Z_{11}+g_2 U^Z_{21}+g_3U^Z_{31})^2}{m^2_{\mathcal{Z}_1}}\right.\crn
&&\left. + \fr{(g_1 U^Z_{12}+g_2 U^Z_{22}+g_3U^Z_{32})^2}{m^2_{\mathcal{Z}_2}}+ \fr{(g_1 U^Z_{13}+g_2 U^Z_{23}+g_3U^Z_{33})^2}{m^2_{\mathcal{Z}_3}}\right],\label{fcncb} \\  \Upsilon^{\prime ij}_R&=&\frac{1}{3}\left[\left( V^*_{qR}\right)^{3i}\left(V_{qR}\right)^{3j} \right]^2\left[\fr{(g_4 U^{Z}_{31})^2}{m^2_{\mathcal{Z}_1}}+\fr{(g_4 U^{Z}_{32})^2}{m^2_{\mathcal{Z}_2}}+\fr{(g_4 U^{Z}_{33})^2}{m^2_{\mathcal{Z}_3}}\right].
\label{FCNC4a}\eea

This effective Lagrangian contributes to mass splittings $\Delta m_M$ between neutral mesons $M^0$-$\bar{M}^0$, where $M$ denotes $B_{d,s}$ or $K$. With the help of the mass matrix elements in \cite{matrixelements}, the mass differences computed from (\ref{fcncb}) and (\ref{FCNC4a}) are  
\bea
\Delta m_K &&=\fr{2}{3}\Re \left\{ \Upsilon'^{12}_L+\Upsilon'^{12}_R \right\} m_Kf_K^2,\label{dmk} \\ \Delta m_{B_{d}}&&=\fr{2}{3}\Re \left\{ \Upsilon'^{13}_L+\Upsilon'^{13}_R \right\} m_{B_d}f_{B_d}^2, \label{dal} \\ \Delta m_{B_{s}}&&=\fr{2}{3}\Re \left\{ \Upsilon'^{23}_L+\Upsilon'^{23}_R \right\} m_{B_s}f_{B_s}^2.\label{dmb} \eea

The total mass differences can be decomposed as   
\bea
\left(\Delta m_{M}\right)_{tot}=\left(\Delta m_{M}\right)_{SM}+\Delta m_{M},
\eea
where the first term comes from the standard model contribution given in \cite{SMFCNC} and
the second term is the new physics contribution as derived in  (\ref{dmk}), (\ref{dal}), or (\ref{dmb}). These predictions are compared to the experimental values  \cite{SMFCNC}. Here, for the neutral kaon mixing we assume that the theory predicts the mass difference within $30\%$ since the potential long-range uncertainties are large. In contrast, the intrinsic theoretical uncertainties for $B_{s,d}$ mass differences are small, assumed to be within $5\%$. In other words, the meson mass differences obey
\bea
0.37044 \times 10^{-2} /ps &&  < \left (\Delta m_K \right)_{tot} < 0.68796 \times 10^{-2} /ps, \\
0.480225/ps && < \left (\Delta m_{B_d} \right)_{tot} < 0.530775 /ps, \\
16.8692/ps &&  < \left (\Delta m_{B_s} \right)_{tot} < 18.6449 /ps. 
\eea
  
\begin{figure}[h]
	\centering
	\begin{tabular}{cc}
		\includegraphics[width=8cm]{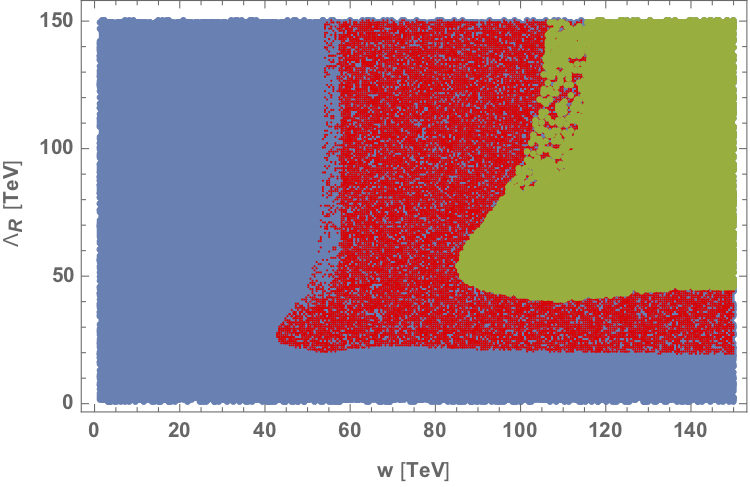}& \includegraphics[width=8cm]{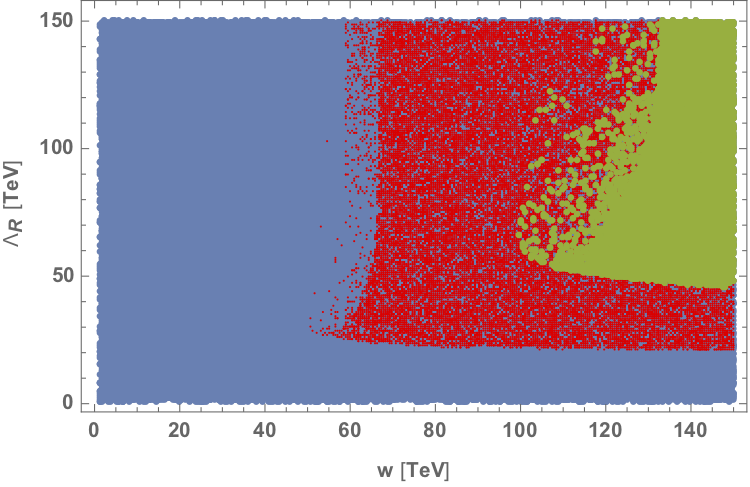} \\
	\end{tabular}
	\caption{Contours of $\Delta m_K$, $\Delta m_{B_s}$, and $\Delta m_{B_d}$ as a function of $(w,\La_R)$ according to $\beta=-\fr{1}{\sqrt{3}}$ (left panel) and $\beta=\fr{1}{\sqrt{3}}$ (right panel). 
		\label{hhh}}
\end{figure}

For a numerical investigation, we  take $w= w_R$, $g_L=g_R$ (i.e. $t_R=1$), $V_{uL}=V_{uR}=I$, and $\La_R,w$ are beyond the weak scale and free to float,  $V_{dL}=V_{dR}=V_{\mathrm{CKM}}$ (i.e., the misalignment in $V_{\mathrm{CKM}}$ tight to the down-quark sector), where the left and right values equal due to the left-right symmetry. With the input parameters $V_{\mathrm{CKM}}, m_{K,B_{s,d}}, f_{K, B_{s,d}}$ given in \cite{Tanabashi:2018oca} and the new neutral gauge boson masses derived by numerical diagonalization of the $M_{3\times 3}$  matrix in Appendix \ref{Appendix}, we make contours of the mass differences, $\Delta m_K$ and $\Delta m_{B_{d,s}}$ in $w$-$\La_R$ plane as Fig. \ref{hhh}. The viable regime (gray) for the kaon mass difference is almost entirely the frame. The red and olive regimes are viable for the mass differences $\Delta m_{B_s}$ and $\Delta m_{B_d}$,
respectively. Combined all the bounds, we obtain  $w>85$ TeV and $\La_R > 54$ TeV  for the model with $\beta=-\fr{1}{\sqrt{3}}$, whereas $w>99$ TeV, $\La_R > 66$ TeV for the model with $\beta=\fr{1}{\sqrt{3}}$. Here the $\beta$ values chosen correspond to the dark matter versions \cite{3331DM}. 

\section{\label{CLFV} Charged LFV}

One of the strongest bounds on the charged LFV is the decay $\mu\to e\ga$. Hence, in this work we study that channel in detail and discuss other charged LFV processes which are potentially troublesome. 

\mathversion{bold}
\subsection{$\mu\to e\gamma$ decay rate}
\mathversion{normal}

We are going to derive an expression for the branching decay ratio of 
 $\mu\to e\gamma$ in the flipped trinification, based upon $SU(3)_C\otimes SU(3)_L\otimes SU(3)_R\otimes U(1)_X$ gauge symmetry, completed by a left-right symmetry of $SU(3)_L$ and $SU(3)_R$ interchange. Similarly to the standard model, the decay $\mu\to e\gamma$ in the present model cannot occur at tree-level, but prevails happening through one-loop diagrams, which are contributed by new Higgs scalars, new gauge bosons, and new leptons.
  
Suppose that the gauge states and the mass eigenstates of the new ($N$) and ordinary charged ($l$) leptons are related as follows 
\bea
N_{a L}=(U^{N}_L)_{a k} N_{kL}^{\prime}, \hs N_{a R}=(U^{N}_R)_{a k} N_{k R}^{\prime},
\hs
 e_{a L}=(U^{l}_L)_{a k}e_{kL}^{\prime}, \hs e_{a R}=(U^{l}_R)_{a k}e_{kR}^{\prime},
\eea
where $U^{N}_{L,R}$ and $U^{l}_{L,R}$ are basis-changing (mixing) matrices and unitary. If the left-right symmetry is not imposed, i.e. $w_L=0$ as in the previous study \cite{3331DM}, $U^{N}_{L,R}$ and $U^{l}_{L,R}$ are not independent, because the mass matrices of $N$ and $l$ are solely generated by the same Yukawa coupling $y_{ab}$ [cf. Eqs. (\ref{maN}) and (\ref{mal})]. It is easily realized that the one-loop diagrams with the mediation of charged gauge $Y^{\pm(q+1)}$ or Higgs $\mathcal{H}^{\pm(q+1)}$ bosons that couple to $l,N$ do not contribute to the decay $\mu\to e \gamma$, since the new leptons do not mix in the basis of charged lepton eigenstates due to the mentioned $m_l,m_N\sim y$. Alternatively, when the left-right symmetry is included, the mass matrices of $l$ and $N$ generally differ due to the $z_{ab}$ coupling contribution, where note that $0\neq w_L\ll w,w_R\sim M$ recognize a left-right asymmetry at the low energy. In this case, the new fields $Y^{\pm(q+1)}$/$\mathcal{H}^{\pm(q+1)}$ and $N$ significantly contribute. That said, the two cases must be taken into account when we parametrize the mixing matrices for numerical investigation, in the following section.

The neutrino mixing matrix is denoted as $U^\nu$, which is a $6\times 6$ unitary matrix, relating the gauge state $X_L\equiv(\nu_L, (\nu_R)^c)^T$ to the mass eigenstate $X'_L$, such as $X_L=U^\nu X'_L$. We write $U^\nu$ in terms of 
\bea
U^\nu=\left(
  \begin{array}{cc}
      U_L & U_A \\
      U_B& U_R  \\
  \end{array}
\right)=\left(
  \begin{array}{c}
      U_L^\nu \\
      U_R^\nu \\
  \end{array}
\right).
\eea
Hence, the Yukawa coupling $x$ can be easily written in terms of diagonal light (called $m_L$) 
and heavy (called $m_R$) mass matrices and the mixing matrices $U_{L,R,A,B}$,
\be
\label{YukawaX}
x=-\frac{\left(U_L^*m_LU_L^\dagger+U_A^*m_RU_A^\dagger\right)}{\sqrt2\Lambda_L}=
-\frac{\left(U_R^*m_RU_R^\dagger+U_B^*m_LU_B^\dagger\right)}{\sqrt2\Lambda_R}.
\ee
Similarly, the Yukawa coupling $y$ can be expressed in terms of the diagonal matrices $m_l, m_N$ that include respective charged and new lepton masses and the mixing matrices $U_{L,R}^{l,N}$,
\be
\label{YukawaY}
y=-\frac{\sqrt2U_L^l m_l\left(U_R^l\right)^\dagger}{u'}=-\frac{\sqrt2U_L^N m_N\left(U_R^N\right)^\dagger}{w}-z\fr{w_Lw_R}{\sqrt{2}Mw}.
\ee 

To derive the decay rate $\mu\to e \gamma$ at one-loop approximation, we necessarily 
calculate the form factors of the relevant one-loop diagrams that contribute to the process. We list 
in Table \ref{TableVertex} the vertexes that are present in the current model and involved in the phenomenon 
of interest. In the table, we denote $i=1\ (2)$ according to either $c_{\xi}$ or $s_{\xi}$ out (in) the brackets, respectively. Previous works of the process $\mu \to e \ga$ have calculated the form
factors by taking into account the contributions of singly-charged gauge bosons ($W$ boson), 
doubly-charged Higgs scalars \cite{Petcov:1976ff, Bilenky:1977du, Cheng:1980tp, Ma:2000xh, Bernabeu:1985na,
Leontaris:1985qc}, as well as singly-charged Higgs scalars for the first time in~\cite{Dinh:2012bp}. In this
paper, we present the results for the form factors of one-loop diagrams with the exchange of 
virtual general charged Higgs scalars and gauge bosons. To our best knowledge, this has not been done so far.

\begin{table}[h]
	\scalebox{1}{
		\begin{tabular}{|c|c|}
			\hline
			Vertex & Coupling \\
			\hline
			$\bar{e}'_{L} X'^c_{L} H^-_i$ & $  Y_{H_i^-}^{L}= -i(U^l_L)^\dagger (x~y) (U^{\nu})^*c_{\xi_5}~(s_{\xi_5})$ 
			 \\
			\hline
			$\bar{e}'_{R} X'_{L} H^-_i$ & $   Y_{H_i^-}^{R}=-i(U^l_{R})^\dagger (y~x) U^{\nu} s_{\xi_5}~(c_{\xi_5})$ \\
			\hline
			$\bar{e}'_{L} N'_{R}  \mathcal{H}_{i}^{-(q+1)}$&$ Y_{\mathcal{H}_i^{-(q+1)}}^L=-i(U^l_L)^\dagger y U^{N}_Rc_{\xi_4}~(s_{\xi_4})$ \\
			\hline
			$\bar{e}'_{R} N'_{L}  \mathcal{H}_{i}^{-(q+1)}$&$ Y_{\mathcal{H}_i^{-(q+1)}}^R=-i(U^l_R)^\dagger y U^{N}_Ls_{\xi_4}~(c_{\xi_4})$ \\
			\hline
			$\left(\bar{e}'_{L}\gamma^\mu \nu'_{L}\right) W_{i\mu}^{-}$ &$\frac{-ig_L}{\sqrt{2}}U_L^{W_{i\mu}^{-}}=\frac{-ig_L}{\sqrt{2}}(U^{l\dagger}_L U^{\nu}_L)c_{\xi}~(s_{\xi})  $ \\
			\hline
			$\left(\bar{e}'_{R}\gamma^\mu \nu'_{R}\right) W_{i\mu}^{-}$ &$\frac{-ig_R}{\sqrt{2}}U_R^{W_{i\mu}^{-}}=\frac{-ig_R }{\sqrt{2}}(U^{l\dagger}_R U^{\nu}_R)s_{\xi}~(c_{\xi})  $ \\			
			\hline
			$\left(\bar{e}'_{L}\gamma^\mu N'_{L}\right) Y_{i\mu}^{-(q+1)}$ &$\frac{-ig_L}{\sqrt{2}}U_L^{Y_{i\mu}^{-(q+1)}}=\frac{-ig_L }{\sqrt{2}}(U^{l\dagger}_L U^{N}_L)c_{\xi_2}~(s_{\xi_2})   $ \\ 			
			\hline
			$\left(\bar{e}'_{R}\gamma^\mu N'_{R}\right) Y_{i\mu}^{-(q+1)}$ &$\frac{-ig_R}{\sqrt{2}}U_R^{Y_{i\mu}^{-(q+1)}}=\frac{-ig_R }{\sqrt{2}}(U^{l\dagger}_R U^{N}_R)s_{\xi_2}~(c_{\xi_2})   $ \\ 			
			\hline
			$\bar{e}'_{L} e'^c_{L} H_{i}^{--}$ &$ Y_{H_i^{--}}^L= -i(U^l_L)^\dagger x (U^{l}_L)^*c_{\xi_6}~(s_{\xi_6}) $\\ 			
			\hline
			$\bar{e}'_{R} e'^c_{R} H_{i}^{--}$ &$ Y_{H_i^{--}}^R=-i(U^l_R)^\dagger x (U^{l}_R)^*s_{\xi_6}~(c_{\xi_6}) $\\ 			
			\hline
	\end{tabular}}
	\caption{Vertexes that contribute to the decay rates $\ell\to \ell'\gamma$. }
	\label{TableVertex}
\end{table}

The effective Lagrangian derived from calculations of the form factors of one-loop 
diagrams for $\mu \to e\ga$ with participation of virtual scalars and gauge bosons in the considering model can be simply expressed as
\begin{equation}
\label{effectiveLag}
\mathcal{L}_{eff}=-4\frac{eG_F}{\sqrt{2}}m_\mu \left( A_R\bar{e}\sigma_{\mu\nu}P_R\mu
+A_L\bar{e}\sigma_{\mu\nu}P_L\mu\right) F^{\mu\nu}+H.c.
\end{equation}
Here $A_{L,R}$ are the form factors:
\begin{eqnarray}
\nonumber
A_R=&&-\sum_{{H^Q},k}\frac{1}{192\sqrt{2}\pi^2G_F M_H^2}\left[\left(Y_H^L\right)_{\mu k}\left(Y_H^L\right)_{e k}^*\times F(Q)
+\frac{m_k}{m_\mu}\left(Y_H^R\right)_{\mu k}\left(Y_H^L\right)_{e k}^*\times 3\times F(r,s_k,Q)\right]\\
&&+\sum_{A_\mu^Q,k}\frac{1}{32\pi^2}\frac{M_w^2}{M_{A_\mu}^2}\left[\left(U_{A_\mu}^L\right)_{\mu k}\left(U_{A_\mu}^L\right)_{e k}^*
G_{\gamma}^Q(\lambda_k)-\left(U_{A_\mu}^R\right)_{\mu k}\left(U_{A_\mu}^L\right)_{e k}^*
\frac{m_k}{m_\mu}R_{\gamma}^Q(\lambda_k)\right],\label{FunAR}
\end{eqnarray}
\begin{eqnarray}
\nonumber
A_L=&&-\sum_{{H^Q},k}\frac{1}{192\sqrt{2}\pi^2G_F M_H^2}\left[\left(Y_H^R\right)_{\mu k}\left(Y_H^R\right)_{e k}^*\times F(Q)
+\frac{m_k}{m_\mu}\left(Y_H^L\right)_{\mu k}\left(Y_H^R\right)_{e k}^*\times 3\times F(r,s_k,Q)\right]\\
&&+\sum_{A_\mu^Q,k}\frac{1}{32\pi^2}\frac{M_w^2}{M_{A_\mu}^2}\frac{g_R^2}{g_L^2}\left[\left(U_{A_\mu}^R\right)_{\mu k}\left(U_{A_\mu}^R\right)_{e k}^*
G_{\gamma}^Q(x)-\left(U_{A_\mu}^L\right)_{\mu k}\left(U_{A_\mu}^R\right)_{e k}^*
\frac{m_k}{m_\mu}R_{\gamma}^Q(\lambda_k)\right],\label{FunAL}
\end{eqnarray}
where $H^Q=H^{+}_{i},H^{++}_{i},\mathcal{H}^{+(q+1)}_{i}$, $A_\mu^Q=W_{i\mu}^+,Y_{i\mu}^{+(q+1)}$ $(i=1,2)$, and
$m_k$ are the masses of associated fermions that along with either $H^Q$ or $A_\mu^Q$ form loops. The functions $F(Q)$, $F(r,s_k,Q)$,
$G_{\gamma}^Q(x)$, and $R_{\gamma}^Q(x)$ appearing in Eqs. (\ref{FunAR}) and (\ref{FunAL}) are defined as
\bea
F(Q)&=&\frac{3}{4}Q-\frac{1}{2},\\
F(r,s_k,Q) &=& Q-\frac{1}{2}-(Q-1)\crn
&&\times \left[\frac{4s_k}{r}+\log(s_k)+(1-\frac{2s_k}{r})\sqrt{1+\frac{4s_k}{r}}
\log\left(\frac{\sqrt{r+4s_k}+\sqrt{r}}{\sqrt{r+4s_k}-\sqrt{r}}\right)\right],\\
G_\gamma^{Q}(x)&=&\frac{(9Q+2)x^2-(12Q-5)x+3Q-1}{4(x-1)^3}-\frac{3}{2}\frac{x^2(Qx-Q+1)}{(x-1)^4}\log(x),
\\
R_\gamma^{Q}(x)&=&-\frac{(2Q-1)x^2+(2Q-1)x-4(Q+1)}{2(x-1)^2}+\frac{3x(Qx-Q-1)}{(x-1)^3}\log(x),
\eea
where we have defined $\lambda_k=m_k^2/M_{A_\mu^Q}^2$, $s_k=m_k^2/M_{H^Q}^2$, $r=-q^2/M_{H^Q}^2$, and that $q=p_2-p_1$ is
transferred momentum.

The branching ratio of $\mu\to e+\gamma$ decay is obtained as \cite{Bilenky:1977du, Cheng:1980tp}  
\begin{equation}
\label{mutoegammarate}
{\rm Br}{(\mu\to e+\gamma)}=384\pi^2(4\pi\alpha_{em})\left(|A_R|^2+|A_L|^2\right),
\end{equation}
where $\alpha_{em}=1/128$ is the fine-structure constant. 

\mathversion{bold}
\subsection{Numerical analysis/discussion: $w_L=0$}
\label{SectionWL0}
\mathversion{normal}

Before performing numerical calcualtions using the branching decay formula obtained in the previous section, let us estimate the magnitudes of relevant VEVs. Among the VEVs introduced, the smallest one could be $\Lambda_L$, which is at eV scale responsible for the neutrino masses, much smaller than the weak scales $u,u'$ satisfying the constraint $u^2+u'^2=(246~{\rm GeV})^2$. Hence, we safely neglect the contributions of $\Lambda_L$. The quark FCNC constraints imply $w,w_R,\La_R \gtrsim \mathcal{O}(50-100)$ TeV, appropriate to the collision bounds \cite{3331DM}, where such VEVs break the flipped trinification to the standard model, significantly greater than the weak scales. Finally, $\Lambda_R$ can take a value, such that (i) $\Lambda_R\gg w,w_R$, (ii) 
$\Lambda_R\sim w,w_R$, or (iii) $\Lambda_R\ll w,w_R$, depending on the symmetry breaking scheme. The viable dark mater scenarios~\cite{3331DM, Huong:2016ybt} prefer the cases (i) and (ii), which will be taken into account.

Due to the condition $\Lambda_R,w,w_R\gg u,u'\gg w_L,\Lambda_L$, the masses of the gauge 
bosons relevant to the process are approximated as 
$m^2_{W_1}\simeq \fr{g^2}{4}(u^2+u'^2)$, $m^2_{W_2}\simeq \frac{g^2}{2}\Lambda^2_R$,
$m^2_{Y_1}\simeq  \frac{g^2}{4}w^2$, and 
$m^2_{Y_2}\simeq \frac{g^2}{4}\left(w^2+w_R^2\right)$, where we have used $g_L=g_R=g$. Note that $W_1$ has the mass identical to the standard model, while $m_{W_2}$ and $m_{Y_{1,2}}$ are large, at TeV scale 
or higher. The masses of relevant new Higgs bosons $H^\pm$, $H^{\pm\pm}$, and $H^{\pm(q+1)}$ depend on unknown parameters present in the scalar potential, which cannot be estimated precisely. However, their masses are all proportional to the new physics scales $\La_R,w,w_R$, which should be large enough in order to escape from detections \cite{Tanabashi:2018oca}. That said, it is reasonable to 
choose the new Higgs masses from hundreds of GeV to few TeV. Particularly, in hierarchical cases the largest masses can be chosen up to 
hundreds of TeV.

Let us parametrize the Yukawa couplings and mixing matrices, involved in the branching ratio $\mu\to e\gamma$ in (\ref{mutoegammarate}), in forms convenient to numerical investigation using the current data. Without lost of generality, we work in the basis of charged lepton mass eigenstates, i.e. $m_l=yu'/\sqrt 2$ is diagonal, implying $U_{L,R}^l=I$, in the same criteria used in the standard model. Besides, the new lepton masses are generated by the same Yukawa matrix $y$, with the relation between both kinds of masses has been given in (\ref{YukawaY}), where $w_L=0$. Hence, the choice of $U_{L,R}^l$ leads to $U_{L,R}^N=I$. Without dependence of basis, the ratio $m^i_l/m^i_N$ $(i=1,2,3)$ is universal for any generation, 
 \be
 \label{memE}
 \frac{m_l^i}{m_N^i}=\frac{u'}{w},
 \ee which imply $m^i_N\gtrsim 50$ MeV, 10 GeV, 170 GeV for $i=1,2,3$, respectively, given that $w/u'\gtrsim 100$. We are interested in the two dark matter versions according to $\beta=\pm1/\sqrt{3}$, where note that $N$ is a standard model singlet for $\beta=-1/\sqrt{3}$, whereas it has an electric charge $q=-1$ for $\beta=1/\sqrt{3}$. The former is always viable in similarity to the case of a light sterile neutrino. However, the latter should be ruled out due to the electroweak precision test, unless the new physics scale is unexpectedly raised, $w/u'\gtrsim 10^5$, so that the lightest new lepton is heavy enough to suppress the dangerous processes, e.g. $Z\rightarrow NN$.  
 
Note that at one-loop approximations, the diagrams with virtual neutral Higgs scalars do not contribute to LFV processes, including $\mu\to e\gamma$ decay, because the interacting vertexes of two leptons with such a neutral scalar do not change flavor (i.e. conserving flavor). The vertex couplings are governed by the magnitudes of diagonal elements of the Yukawa matrix $y$ as well as a mixing factor among neutral scalars. These vertexes are also constrained by the current experiments through the channels of the standard model like Higgs decay into two leptons $h\to \ell\bar{\ell'}$. According to \cite{Tanabashi:2018oca}, $h\to \tau\bar{\tau}$ has been observed at a quite high precision, while 
$h\to \mu\bar{\mu}$ is likely observed, but at large uncertainty, and the branching decay $h\to e\bar{e}$ can only be set by an upper limit, $\mathrm{Br}(h\to e\bar{e})<1.9\times 10^{-3}$. All these agree to the strengths of $h \ell\ell$ interactions, set by the corresponding lepton masses. Due to the mixing, $h$ can decay into light $N$'s, but the rate is highly suppressed  by $(u,u')^2/(w,w_R,\La_R)^2\ll 1$. The light $N$'s are undetectable due to weak interaction strengths. However, since no constraint has been made to their masses, they can take any values consistent to the scenario of interest.

Our study is interested in a model whose new physics scale is not too high, thus presenting rich physical phenomena at the current and future experiments. Let 
us fix the benchmark values for $u'$ and $w$, based upon the relation (\ref{memE}) and the others. Close to the standard model, we choose $u'$ near its maximum, $u'\simeq 246$ GeV. Another advantage of this choice leads to the smallness of $u$, thus to have a significant reduction of $\La_R$ satisfying the required condition for the seesaw mechanism, $u^2/\Lambda_R\sim \rm{eV}$. Choosing the lower bound for the heaviest new lepton to be $50$, 200, 500 GeV, one has $w\geq 7,\ 28,\ 70$ TeV, respectively. In the seesaw mechanism, note that $U_L$ and $U_R$ diagonalize the mass matrix $M_\nu$ and $M_R$, respectively. Thus, $U_L$ and $U_R$ generally differ. The observed neutrino masses imply the sizes of $M_L$ and $M_D^T M_R^{-1}M_D$ at eV.  The smallness of $M_L$ is 
ensured by small $\Lambda_L\sim\mathrm{eV}$, while the 
magnitude of $M_D^TM_R^{-1}M_D$ depends on the correlation between $y$, $u$, and $M_R$. Since $y$ has been fixed before 
$\mathrm{diag}(y)=(3\times 10^{-6},6\times 10^{-4}, 10^{-2})$, the lower
bound set for the right-handed neutrino masses $M_R$ is 10 and 1000 TeV according to
$u=0.1$ and 1 GeV, respectively.

Summarizing all, the parmetrization for numerical investigation is now performed. First of 
all, $U_L$ coincides with the Pontecorvo-Maki-Nakagawa-Sakata matrix $U_{\mathrm{PMNS}}$, determined with a high accuracy by the oscillation experiments, except for the Dirac CP violation phase (where the Majorana CP violation
phases are neither determined nor contributing to the process). The $3\times 3$ unitary
matrix $U_R$ is parametrized in the same way as $U_{\mathrm{PMNS}}$, but its angles and phase are freely chosen in 
the calculation. Lastly, $M_R$ can be 
calculated using the relation  $M_R=U^*_R M_R^{diag} U^\dagger_R$, where 
$M_R^{diag}$ has a diagonal form of the heavy neutrino masses. The Yukawa
matrix $x$ is derived from $M_R$ as
$x=-U^*_R M_R^{diag} U^\dagger_R/(\sqrt 2 \Lambda_R)$. In the following, we will present the results of numerical calculations for the case where the involved parameters are chosen as $u=0.1$ GeV, $w=10$ TeV,
$w_R=20$ TeV, ${\rm{diag}} (M_R^{diag})=(10,20,30)$ TeV 
and $U_R(\theta_{12}',\theta_{13}',\theta_{23}',\delta')=U_R(\pi/4,\pi/4,0,0)$. Note that the choice $u=0.1$ 
GeV is in order to conserve  
the condition $u'\simeq 246$ GeV, whose important implications have been discussed before. The other quantities such as 
$\Lambda_L$, $w_L$, and light neutrino masses are neglected due to the small effects for the process.

Although $Y_i,\mathcal{H}_i$ are listed in Table \ref{TableVertex}, it is realized that 
their vertexes do not contribute to the $\mu\to e\gamma$ branching 
ratio at the one-loop level, as mentioned. The reason is similar to the case of vertexes of neutral Higgs scalars, which conserve lepton flavors. Indeed,
 all of the matrices relevant to them, such as $U_{R,L}^l$, $U_{R,L}^N$ and $y$, are diagonal.

\begin{figure}[h]
\begin{center}
\includegraphics[width=11cm]{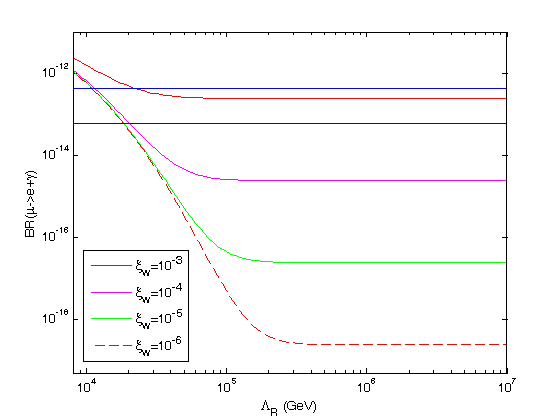}
  \caption{The branching ratio $\mathrm{Br}(\mu\to e\gamma)$ governed by intermediate $W_{1,2}^\pm$ gauge bosons, which is given as a function of $\Lambda_R$ for the selected values of their mixing angle $\xi_w$. The upper and lower blue lines correspond to the MEG current bound and near-future sensitivity limit.}
\label{GaugeGaR}
 \end{center}
\end{figure}

\begin{figure}[h]
\begin{center}
\includegraphics[width=11cm]{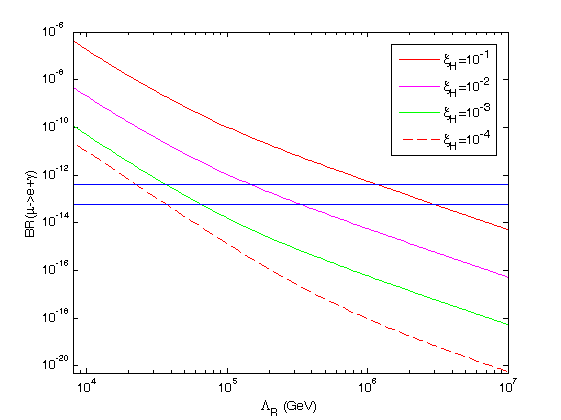}
  \caption{The branching ratio $\mathrm{Br}(\mu\to e\gamma)$ governed by intermediate Higgs bosons $H_{1,2}^\pm$ and $H_{1,2}^{\pm\pm}$, which is given as a function of $\Lambda_R$ for the selected values of the mixing angle $\xi_H$. Here, we have set $M_H=10$ TeV as a common mass for all $H_{1,2}^\pm,H_{1,2}^{\pm\pm}$ and supposed that the pairs $H_1^\pm$-$H_2^\pm$ and $H_1^{\pm\pm}$-$H_2^{\pm\pm}$ have the same mixing angle ($\xi_H$).}
\label{HiggsGaR}
 \end{center}
\end{figure}

\begin{figure}[h]
\begin{center}
\begin{tabular}{cc}
\includegraphics[width=7.5cm,height=6.5cm]{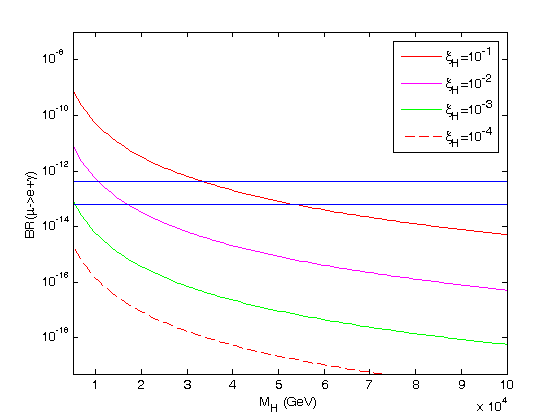} &
\includegraphics[width=7.5cm,height=6.5cm]{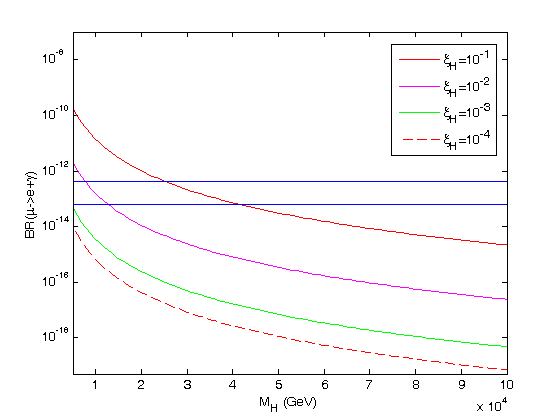}
\end{tabular}
\caption{The branching ratio $\mathrm{Br}(\mu\to e\gamma)$ governed by either virtual singly-charged scalars (left panel) or virtual doubly-charged scalars (right panel), all given as a function of the Higgs common mass $M_H$ (all the Higgs bosons presumably have the same mass) for different values of their mixing angle $\xi_H$ with fixed $\Lambda_R=100$ TeV.}
\label{QDHiggsMassSingDoubChage}
\end{center}
\end{figure}

\begin{figure}[h]
\begin{center}
\includegraphics[width=11cm]{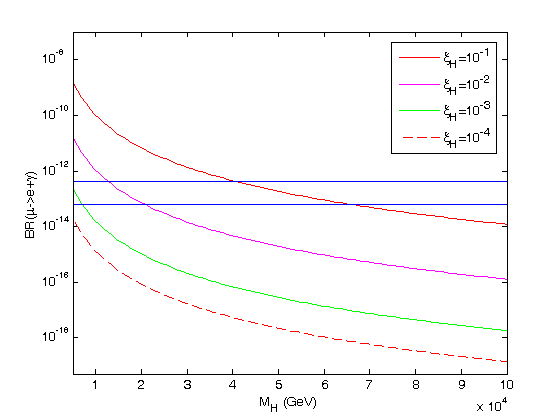}
  \caption{Dependence of the branching ratio $\rm {Br}(\mu\to e\gamma)$ as governed by the exchanges of virtual singly and doubly charged scalars, given as a function of the Higgs common mass $M_H$ for different cases of the mixing angle $\xi_H$, where $\Lambda_R=100$ TeV is fixed.}
\label{HiggsMass}
 \end{center}
\end{figure}

In Figs. \ref{GaugeGaR}, \ref{HiggsGaR}, \ref{QDHiggsMassSingDoubChage} and \ref{HiggsMass}, we 
respectively show the dependence of the branching ratio $\rm {Br}(\mu\to e\gamma)$ on the relevant parameters in this kind of the model. To produce the results, we have separately considered the contributions to the decay rate corresponding to the exchanges of the virtual gauge boson and charged Higgs scalar, respectively. As have been introduced above and expressed in detail 
from Eq. (\ref{effectiveLag}) to Eq. (\ref{mutoegammarate}), in the model under consideration
the $\mu\to e\gamma$ branching ratio depends complicatedly on many parameters, where most of them are 
unknown. Moreover, the variation of a parameter might change the contribution of the involving channel to 
few orders, e.g. the mixing angles between the left and right sectors $\xi$, see the figures for details.  
Therefore, presenting individual contributions would provide more information and better understanding
about the phenomenon. We also suppose that heavy Higgs $H^{\pm}_i$, $H^{\pm\pm}_i$ $(i=1,2)$ possess equivalent 
masses, commonly called $M_H$. Additionally, the mixing angles between $H_1^\pm$ and $H_2^\pm$ as well as beween $H_1^{\pm\pm}$ and $H_2^{\pm\pm}$ are equally taken and denoted as $\xi_H$.

Fig. \ref{GaugeGaR} describes the dependence of $\mu\to e\gamma$ upon the 
diagrams that exchange virtual gauge bosons $W_i$ (i=1,2), given as a function of $\Lambda_R$.
The choice of ${\rm{diag}} (M_R^{diag})=(10,20,30)$ TeV and $U_{L,R}$ has canceled out 
the dependence of $U_{mix}$ on $\Lambda_R$. Moreover, $W_1$ is the standard model $W$ boson whose 
mass is fixed as $M_{W_1}=80$ GeV. Therefore, the branching ratio lines shown
in Fig. \ref{GaugeGaR} are depicted as a function of the new boson
mass $M_{W_2}\sim \Lambda_R$ and mixing angle $\xi_w$. For each value of $\xi_w$, the 
branching ratio goes down due to the dominant contribution of $W_2$ to a constant value, as increasing $\Lambda_R$. The constant line is preserved by a constant 
contribution of $W_1$. Using the MEG current bound on the $\mu\to e\gamma$ decay, one roughly estimates 
the lower bound $\Lambda_R\geq 12,\ 10$ TeV for $\xi_w=10^{-3}$ and $\xi_w\leq 10^{-4}$, 
respectively. The strong dependence of the branching ratio on the mixing angle $\xi_w$, which separates about  
two order between two successive lines for the range of large $\Lambda_R$, suggests the domination of the interference terms in $A_{L,R}$ 
[cf. Eqs. (\ref{FunAR}) and (\ref{FunAL})]. Indeed, the interference terms are proportional to 
$\frac{m_k}{m_\mu}\sin\xi_w\cos\xi_w\approx \frac{m_k}{m_\mu}\xi_w $. Thus, the branching ratio 
is proportional to $\frac{m_k^2}{m_\mu^2}\xi_w^2\sim \xi_w^2$, which is consistent to the observation from the figure, whereas the other terms are either proportional to $\cos\xi_w^2\simeq 1$ 
or suppressed by a factor $\xi_w^2$. It is figured out that the dominant interference terms 
are provided by the factor $\frac{m_k}{m_\mu}\sim 10^5$, for instance, for the 
case of heavy neutrino mass $m_k\sim 10^4$ GeV. Similarly, we have the 
same domination of the interference terms in Figs. \ref{HiggsGaR}, \ref{QDHiggsMassSingDoubChage} 
and \ref{HiggsMass}.

The branching ratio in Fig. \ref{HiggsGaR} is a monotonically decreasing function of $\Lambda_R$, which enters the decay rate through the interaction 
vertexes, which have strengths depending on the elements of the Yukawa coupling matrix $x\sim 1/\Lambda_R$.
Behavior of the branching ratio is considered for different values of the mixing angle
$\xi_H$. The figure implies that consistent with the current MEG upper bound, 
$\rm {Br}(\mu\to e\gamma)<4.2\times 10^{-13}$, the lower limits are $\Lambda_R\geq 1100,\ 105,\
13,\ 10.2$ TeV according to $\xi_H=10^{-1},\ 10^{-2},\ 10^{-3}, 10^{-4}$,
respectively. Whilst the sensitivity of the future MEG are possible to probe $\mu\to e\gamma$ signal, provided 
that $\Lambda_R\leq 1300,\ 120,\ 15.5,\ 13$ TeV for $\xi_H=10^{-1},\ 10^{-2},\ 10^{-3},\ 10^{-4},$ respectively.

In the next two figures, we demonstrate the dependence of the branching ratio as a single 
variable function of the new Higgs mass $M_H$, where $\Lambda_R$ is fixed as $100$ TeV.  
As we see from Figs. \ref{QDHiggsMassSingDoubChage} and \ref{HiggsMass}, the smaller 
the mixing angle $\xi_H$ is, the smaller the lower bound is set for the heavy Higgs masses. If the contributions to the diagrams include only the virtual 
singly charged scalar (Fig. \ref{QDHiggsMassSingDoubChage},
left panel), the lower bound for the scalar masses reduces from $M_H=53$ TeV at 
$\xi=10^{-1}$ down to $M_H< 10$ TeV at $\xi=10^{-4}$. We get almost the same limits for the case with doubly charged scalar exchanges (Fig. \ref{QDHiggsMassSingDoubChage},
right panel).

\mathversion{bold}
\subsection{Numerical analysis/discussion: $w_L\neq 0$}

\mathversion{normal}

The flipped trinification discriminates from the minimal left-right symmetric model especially in the extended particle sectors, governed by the new gauge symmetry. Part of them produces the interesting quark FCNCs, as studied above. In this section, we argue that the presence of other part of them gives novel contributions to the charged LFV. It is stressed that such LFV processes e.g. $\mu \to e\gamma$ can be altered in the case of non-vanishing $w_L$. Although $w_L$ is constrained to be much smaller than $M,\Lambda_R,w_R,w$ as well as not modifying the results discussed in the previous section related to $H_i$ and $W_{i}$, the non-vanishing $w_L$ causes the mass matrices of ordinary charged and new leptons to be not simultaneously diagonalized. This provides the new sources of the LFV, which involve the $(q+1)$-charged Higgs and gauge bosons ($\mathcal{H}_{1,2}$, $Y_{1,2}$) as well as the new leptons ($N$) in the loops for $\mu \to e\gamma$, which is a new feature of the model.  

In the basis of ordinary charged lepton mass eigenstates, the mentioned, new lepton mass matrix can 
be expressed as   
\begin{equation}
M_N=M_N^0+\Delta M_N,
\end{equation}
where
\bea
&&M_N^0=-\frac{y_{ab}}{\sqrt{2}}w=\left(
  \begin{array}{ccc}
      m_{N1}^0 & 0 & 0 \\
      0& m_{N2}^0 & 0  \\
      0& 0 & m_{N3}^0 \\
  \end{array}
\right),\\
&&\Delta M_N=-\frac{z_{ab}}{2M}w_L w_R=\left(
  \begin{array}{ccc}
      a_1 & b_1 & b_2 \\
      B_1& a_2 & b_3  \\
      B_2& B_3 & a_3 \\
  \end{array}
\right).
\eea Here $||y||\sim ||M_l||/u'\sim 10^{-3}$--$10^{-2}$ is constrained by the ordinary charged lepon masses and small. The coupling matrix $z$ is generic and maybe sizable, but it generally obeys $||z||\sim 1\ll ||y||(w/w_L)$, provided that $w/w_L\gtrsim 10^{3}$. For instance, if $w=10$ TeV, one takes $w_L\lesssim 10$~GeV. This leads to $||\Delta M_N||\ll ||M^0_N||$, which is also expected due to the contribution of the effective interactions. 
To find the mixing matrix, we dialgonalize pertubatively the squared mass matrix, $MM^{\dagger}$,
while taking into account $\Delta M_N$ as a sub-dominant contribution comparing to $M_N^0$. The final result is
\bea
U\approx\left(
  \begin{array}{ccc}
      1 & \frac{b_1^*m_{N1}^0+B_{1}m_{N2}^0}{(m_{N2}^0)^2-(m_{N1}^0)^2} & \frac{b_2^*m_{N1}^0+B_{2}m_{N3}^0}{(m_{N3}^0)^2-(m_1^0)^2} \\
      -\frac{b_1m_{N1}^0+B_{1}^*m_{N2}^0}{(m_{N2}^0)^2-(m_{N1}^0)^2}& 1 & \frac{b_3^*m_{N2}^0+B_{3}m_{N3}^0}{(m_{N3}^0)^2-(m_2^0)^2}  \\
      -\frac{b_2m_{N1}^0+B_{2}^*m_{N3}^0}{(m_{N3}^0)^2-(m_1^0)^2}& -\frac{b_3m_{N2}^0+B_{3}^*m_{N3}^0}{(m_{N3}^0)^2-(m_2^0)^2}  & 1 \\
  \end{array}
\right).
\eea

\begin{figure}[h]
\begin{center}
\includegraphics[width=11cm]{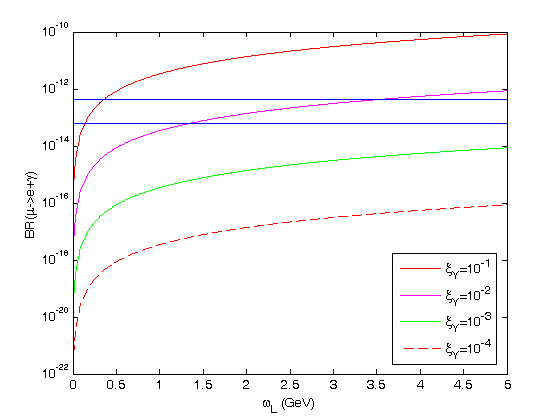}
  \caption{Dependence of the branching ratio $\mathrm{Br}(\mu\to e\gamma)$, governed by the virtual $Y_{1,2}^{\pm(q+1)}$ gauge boson exchanges, on $w_L$ for different values of the mixing angle $\xi_Y$. The upper and lower lines correspond to the MEG current bound and the near future sensitivity limit.}
\label{GaugeYWL}
 \end{center}
\end{figure}

\begin{figure}[h]
\begin{center}
\includegraphics[width=11cm]{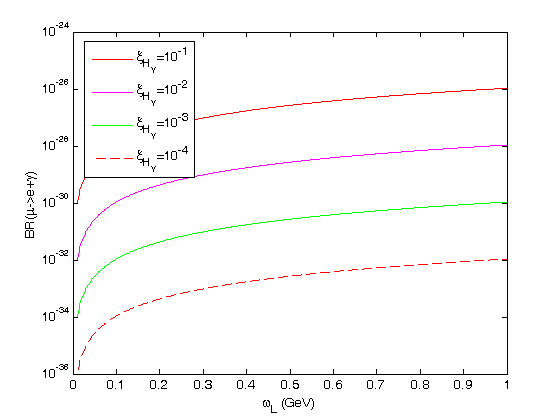}
  \caption{Dependence of the branching ratio $\mathrm{Br}(\mu\to e\gamma)$, governed by the virtual charged Higgs $\mathcal{H}^{\pm(q+1)}_{1,2}$ exchanges, as a function of $w_L$ for different values of the mixing angle $\xi_{H_Y}$, where we fixed the Higgs masses $M_H=10$ TeV.}
\label{HiggsYWL}
 \end{center}
\end{figure}
For brevity, in numerical calculation we assume $z_{ab}$ as a real symmetric matrix with $z_{12}=z_{13}=z_{23}=1$, which 
means that the new lepton mass matrix is invariant under the charge-conjugation and parity transformations. This choice leads to an approximation, $U_L^E\simeq U_R^E\simeq U$. We fix $M=100$ TeV, $w=30$ TeV, and $w_L$ appropriately ranging from an infinitesimal value to a few GeVs. All the remaining parameters take the same values as in 
the previous subsection. 

In Fig. \ref{GaugeYWL}, we depict the dependence of the branching ratio $\mathrm{Br}(\mu\to e\gamma)$, contributed by the exchanges of virtual $Y_{1,2}^{\pm(q+1)}$ gauge bosons, in terms of $w_L$ for several values of the mixing angle $\xi_Y$. With the set of the parameters used in the numerical calculation, the branching ratio of 
$\mu\to e\gamma$ is within the sensitivities of the current and near-future experiments. The upper bounds
$w_L=0.35,\ 3.53$ GeV are set for $\xi_Y=10^{-1},\ 10^{-2}$, respectively. While the next MEG upgrade might probe the decay signal if $w_L\geq 0.13,\ 1.34$ GeV corresponding to $\xi_Y=10^{-1},\ 10^{-2}$, respectively. 
 
The contributions to the decay $\mu\to e\gamma$ by virtual charged Higgs $\mathcal{H}^{\pm(q+1)}_{1,2}$ exchanges are extremely small,
comparing to those by $Y_{1,2}^{\pm(q+1)}$ gauge bosons, if one uses the same values of the model's parameters 
involved in the process. The branching ratios shown in Fig. \ref{HiggsYWL} are smaller than the gauge ones by 14 orders of magnitude,
which are about 13 orders of magnitude below the future MEG sensitivity. It is not hard to see that the 
branching ratios are strongly suppressed by the ordinary charged-lepton Yukawa couplings $y^4$, where the biggest element is only
$y_3\sim 10^{-2}$, which are much smaller than the gauge contribution.

\subsection{Other charged LFV processes} 

In this model, the charged LFV processes such as $\mu \to 3e$ and $\tau \to 3\mu(3e)$ can exist at the tree-level, 
exchanged by the charged Higgs $H^{\pm\pm}_{1,2}$. The $\mu\to3e$ branching ratio in the present scheme with a low scale of new physics of order $10$--$100$ TeV is expected to be in the sensitive 
ranges of the current and near-future experiments. Present upper bound on branching ratios of $\tau \to 3\mu(3e)$ are
in the order of $10^{-8}$ \cite{Tanabashi:2018oca}, which are five orders less stringent than those of $\mu \to 3e$ decay 
at $10^{-12}$ \cite{Tanabashi:2018oca}. Moreover, the $\mu \to 3e$ experiment at Paul Scherrer Institute (PSI) is 
expected to determine signal of ${\rm {Br}}(\mu \to 3e)\geq 10^{-15}$ and its upgrade is sensitive to the $\mu \to 3e$ branching ratio not smaller than $10^{-16}$ \cite{Blondel:2013ia}. Therefore, we need consider only the interested process of $\mu \to 3e$ decay in this research.

It is easily verified that, in contrast to the previously mentioned processes the charged-LFV neutral-Higgs decays e.g. $h\to \mu\tau$ receives only one-loop contributions. On the theoretical side, they 
are strictly suppressed by the heavy particle masses and the loop factor~$1/16\pi^2$. It is easily proved that such processes satisfy all the current bounds with the chosen 
parameter regime, since such experimental bounds are 
less tight~\cite{Tanabashi:2018oca}.

Using the relevant LFV vertexes given in Table \ref{TableVertex}, while keeping in mind that the doubly charged Higgs bosons that dominantly contribute to the $\mu \to 3e$ decay have the transferred 
momenta much smaller than their masses, one can write down the effective Lagrangian as    
\begin{eqnarray}
\mathcal{L}_{eff}(\mu\rightarrow 3e)&=&g_{LS}^{LL}\left(\bar{{e^c_L}}\mu_L\right)\left(\bar{{e^c_L}}e_L\right)
+g_{RS}^{RR}\left(\bar{{e^c_R}}\mu_R\right)\left(\bar{{e^c_R}}e_R\right)\nonumber \\
 &+&g_{LS}^{LR}\left(\bar{{e^c_L}}\mu_L\right)\left(\bar{{e^c_R}}e_R\right)
+g_{RS}^{RL}\left(\bar{{e^c_R}}\mu_R\right)\left(\bar{{e^c_L}}e_L\right).
\end{eqnarray}
Here, we denote $M_{H_i}$ $(i=1,2)$ to be the masses of doubly charged Higgs bosons and 
\begin{eqnarray}
g_{LS}^{LL}&=&-\sum_{i=1}^2\frac{2}{\left(M_{H_i^{}}\right)^2}\left(y_{H_i}^L\right)_{e\mu}\left(y_{H_i}^L\right)_{ee}
,~~
g_{RS}^{RR}=-\sum_{i=1}^2\frac{2}{\left(M_{H_i^{}}\right)^2}\left(y_{H_i}^R\right)_{e\mu}\left(y_{H_i}^R\right)_{ee},\\
g_{LS}^{LR}&=&-\sum_{i=1}^2\frac{1}{\left(M_{H_i^{}}\right)^2}\left(y_{H_i}^L\right)_{e\mu}\left(y_{H_i}^R\right)_{ee},~~
g_{RS}^{RL}=-\sum_{i=1}^2\frac{1}{\left(M_{H_i^{}}\right)^2}\left(y_{H_i}^R\right)_{e\mu}\left(y_{H_i}^L\right)_{ee}.
\end{eqnarray}
The branching ratio is straightforwardly obtained as \cite{Goto:2010sn}
\begin{equation}
{\rm {Br}}(\mu\rightarrow 3e)=\frac{1}{32G_F^2}\left(|g_{LS}^{LL}|^2+|g_{RS}^{RR}|^2+|g_{LS}^{LR}|^2+|g_{RS}^{RL}|^2\right),
\label{BrMu3e}
\end{equation}
where $G_F=1.166\times 10^{-5}{\rm {GeV^2}}$ is the Fermi coupling constant.

For numerical evaluation, without loss of generality, we assume that both of the doubly charged Higgs bosons have the same
mass, thus $M_{H_i}=M_H$ $(i=1,2)$. The mixing angle, $\xi_H$, between $H_1^{++}$ and $H_2^{++}$ is not 
necessary to be considered for the currently interested process. Because of the $\xi_H$ smallness $(\xi_H\ll 1)$, the dominated contributions come 
from the terms involving $2\left(y_{H_1}^L\right)_{e\mu}\left(y_{H_1}^L\right)_{ee}/\left(M_{H_1^{}}\right)^2$ and 
$2\left(y_{H_2}^R\right)_{e\mu}\left(y_{H_2}^R\right)_{ee}/\left(M_{H_2^{}}\right)^2$, which are easily realized as proportional to 
$\cos{\xi_H}^2\simeq 1$, whereas the others are suppressed by a factor either $\sin{\xi_H}$ or $\sin{\xi_H}^2$. In the 
following discussion, we take $\xi_H=0.1$. The branching ratio expressed in Eq. (\ref{BrMu3e}) is, in fact, inversely proportional to $M_H^4\Lambda_R^4$, because it is proportional to $Y_{H_{1,2}}^{R,L}\sim x,$ where $x=-U^*_R M_R^{diag} U^\dagger_R/(\sqrt 2 \Lambda_R)$, $U_R=U_R(\theta_{12}',\theta_{13}',\theta_{23}',\delta')$, and $g_{L(R)S}^{L(R)L(R)}\sim 1/M_H^2$. Therefore, 
the ratio will be strongly suppressed in the large ranges of $\Lambda_R$ 
and $M_H$. Taking $M_R^{diag}=(10,20,30)$ TeV as chosen before, while varying $\theta_{12}',\theta_{13}',\theta_{23}'$
in the range of $[0,\pi/2]$ and $\delta'$ in $[0,2\pi]$, one gets a bound
\begin{equation}
0\leq{\rm {Br}}(\mu\rightarrow 3e)\leq 1.64\times 10^6\left[\frac{1{\rm{TeV}}}{M_H}\right]^4\left[\frac{1{\rm{TeV}}}{\Lambda_R}\right]^4.
\end{equation}
\begin{figure}[h]
\begin{center}
\includegraphics[width=11cm]{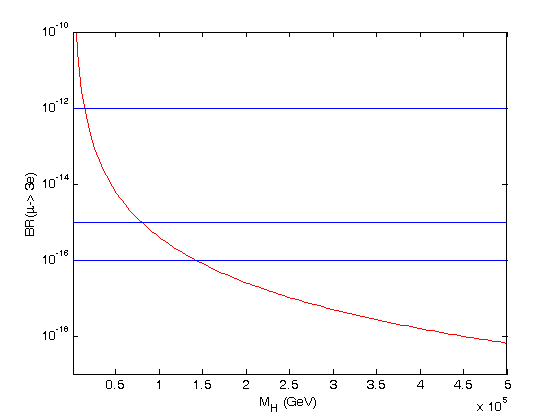}
  \caption{Branching ratio $\mathrm{Br}(\mu\to 3e)$ as a function of doubly 
  charged Higgs masses. The three blue lines, ${\rm {Br}}(\mu\to 3e)=10^{-12};~10^{-15};~10^{-16}$, correspond to 
  the current experimental upper bound, the sensitivities of PSI and PSI upgraded experiments, respectively.} 
\label{BrMu3eFig}
 \end{center}
\end{figure}

As a specific case, taking $\theta_{12}'=\theta_{13}'=\theta_{23}'=\pi/4$ and $\delta'=0$, Fig. \ref{BrMu3eFig}
describes the behavior of the $\mu \to 3e$ branching ratio as a function of the doubly charged Higgs masses. The figure reveals a line of monotonically decreasing function as increasing of $M_H$, which is consistent to the fact 
that the branching ratio is inversely proportional to $M_H^4$, mentioned above. The lower bounds of the 
doubly charged Higgs masses corresponding to the current limit, PSI experiment and 
its upgraded sensitivities are $14,~79,~143$ TeV, respectively. Thus we apparently conclude that the 
future PSI experiment is more sensitive to the new physics of the considering model than the Mu to E Gamma 
experiment (MEG), which gives a lower bound $M_H=53$ TeV for the case $\xi_H=0.1$.  
  
\section{Conclusion \label{Con}}

When a gauge symmetry is flipped, it leads to a deeper structure that defines a more fundamental theory. For instance, 
$SU(2)_L$ flipped yields electroweak unification; $SU(5)$ flipped defines a seesaw mechanism; $SO(10)$ flipped leads to $E_6$ and promising superstring theories. In this work, we have addressed such a nontrivial task, the flipped trinification and its novel consequences. First of all, a trinification flipped unifies both the 3-3-1 and left-right symmetries. Consequently, this flipped trinification resolves the generation number and the weak parity violation. Additionally, it generates neutrino masses and dark matter naturally via the gauge symmetry.

An important feature of the flipped trinification is that it presents the flavor changing currents in both quark and lepton sectors. We have probed that the quark FCNCs bound the new physics scale to be at or beyond several tens of TeV via the neutral meson mixings, $B^0_{d,s}$-$\bar{B}^0_{d,s}$. The charged LFV via the decay $ \mu \to e \ga$ yield mostly the same bound, whereas the other processes such as $\tau\to 3\mu(3e)$ and $h\to \mu \tau$ are easily experimentally satisfied. The process $\mu\to 3e$ receives tree level contributions by the doubly charged Higgs bosons and presents the same limit on the new physics as the meson mixing and $\mu \to e \ga$ do. 

All the results indicate that the trinification is possibly flipped at tens of TeV. Additionally, the contributions of the new particles other than the left-right symmetric model are important to set the charged LFV and quark FCNC observables, which can be used to prove or rule out this proposal.

\section*{Acknowledgements}

This research is funded by Vietnam National Foundation for Science and Technology Development under 
grant number 103.01-2014.89. 

\appendix

\section{Neutral gauge boson mass matrices \label{Appendix}}
For convenience in reading, in this appendix we supply the full neutral gauge boson mass matrix as well as the $3\times 3$ mass matrix of new neutral gauge bosons. 

After the symmetry breaking, the neutral gauge bosons $(A_{3L}, A_{3R}, A_{8L}, A_{8R}, B)$ in such order possess a mass matrix 
\bea
\frac{g_L^2}{4}\left(
\begin{array}{ccccc}
	u_1^2+u_2^2+4 \Lambda_L^2 & -t_R \left(u_1^2+u_2^2\right) & \frac{u_1^2-u_2^2+4 \Lambda _L^2}{\sqrt{3}} & \frac{t_R \left(u_2^2-u_1^2\right)}{\sqrt{3}} & m_{15} \\
	-t_R \left(u_1^2+u_2^2\right) & t_R^2 \left(u_1^2+u_2^2+4 \Lambda _R^2\right) & \frac{t_R \left(u_2^2-u_1^2\right)}{\sqrt{3}} & \frac{t_R^2 \left(u_1^2-u_2^2+4 \Lambda_R^2\right)}{\sqrt{3}} & m_{25} \\
	\frac{u_1^2-u_2^2+4 \Lambda _L^2}{\sqrt{3}} & \frac{t_R \left(u_2^2-u_1^2\right)}{\sqrt{3}} & m_{33} & -\frac{1}{3} t_R \left(u_1^2+u_2^2+4 w^2\right) & m_{35} \\
	\frac{t_R \left(u_2^2-u_1^2\right)}{\sqrt{3}} & \frac{t_R^2 \left(u_1^2-u_2^2+4 \Lambda_R^2\right)}{\sqrt{3}} & -\frac{1}{3} t_R \left(u_1^2+u_2^2+4 w^2\right) & m_{44} & m_{45} \\
	m_{15}&m_{25} &m_{35}&m_{45}&m_{55}\\
\end{array}
\right),\nn
\eea
where we define for short  
\bea
&& m_{15}=-\frac{4t_X (3 + \sqrt{3} \beta) \Lambda_L^2}{3} ,\hs
m_{25}=-\frac{4 t_R t_X \sqrt{3}(\sqrt{3} +\beta)\Lambda_R^2}{3}, \\ 
&& m_{35}=-\frac{4 t_X (w_L^2 \beta +(\sqrt{3}+\beta)\Lambda_L^2)}{3},\hs m_{45}=-\frac{4 t_R t_X (w_R^2 \beta +(\sqrt{3}+ \beta)\Lambda_{R}^2)}{3}, \\  
&& m_{55} = \frac{4t_X^2 (\beta^2(w_L^2+w_R^2)+(\sqrt{3}+\beta)^2(\Lambda_L+\Lambda_R^2))}{3},\\ 
&& m_{33}=\frac{u_1^2+u_2^2+4(w^2+w_L^2+\Lambda_L^2)}{3}, \hs m_{44}=\frac{t_R^2(u_1^2+u_2^2+4(w^2+w_R^2+\Lambda_R^2))}{3}
\eea

Changing to the new basis $(A,Z_L,Z'_L,Z_R,Z'_R)$, we obtain $A,Z_L$ identical to the standard model photon and $Z$ boson respectively, which are light and decoupled. The remaining fields $(Z'_L,Z_R,Z'_R)$ are new and mix via a $3\times 3$ mass matrix,
\bea
M_{3 \times 3}=\frac{g_L^2}{4}\left(
\begin{array}{ccc} m_{11}^\prime & m_{12}^\prime & m_{13}^\prime \\
	m_{21}^\prime & m_{22}^\prime & m_{23}^\prime \\
	m_{13}^\prime & m_{23}^\prime & m_{33}^\prime
\end{array}
\right),
\eea
where $m_{ij}$ are defined as
\bea
m_{11}^\prime&& =\fr{4 c_W^2 w^2}{3(c_W^2-s_W^2 \beta^2)}, m_{12}^\prime= \fr{4t_R w^2 c_W s_W^2}{3(c_W^2-s_W^2 \beta^2)\sqrt{t_R^2-(1+t_R^2)(1+\beta^2)s_W^2}\sqrt{t_R^2\left(1+\fr{\beta^2 s_W^2}{t_R^2-(1+t_R^2)(1+\beta^2)s_W^2} \right)}},\nonumber \\
m_{13}^\prime && =\fr{4\sqrt{2}t_R^2w^2 c_W}{3\sqrt{-1+t_R^2-(1+t_R^2)\left(\beta^2-(1+\beta^2)c_{2W} \right)}\left(1+\fr{\beta^2 s_W^2}{t_R^2-(1+t_R^2)(1+\beta^2)s_W^2} \right)\left(1+\fr{(1+\beta^2) s_W^2}{t_R^2-(1+t_R^2)(1+\beta^2)s_W^2} \right)}, \nonumber \\
m_{22}^\prime && =\fr{4 \left(3t_R^4\La_R^2-6t_R^4(1+\beta^2)\La_R^2 s_W^2+\left(\beta^2 w^2+3t_R^4(1+\beta^2)^2 \La_R^2 \right)s_W^4 \right)}{3\left((1+\beta^2)s_W^2-1\right)\left(-t_R^2+(1+t_R^2(1+\beta^2))s_W^2\right)},\nonumber \\ m_{23}^\prime &&= \fr{-4t_R^2w^2\beta(1+\beta^2)-4t_R^4(\beta w^2+(\sqrt{3}-3 \beta)\La_R^2)(1+\beta^2-c_W^{-2})-4\sqrt{3}t_R^6 \La_R^2(1+\beta^2-c_W^{-2})^2}{3\left(1+t_R^2(1+\beta^2)-t_R^2 c_W^{-2}\right)\left((1+t_R^2)(1+\beta^2)-t_R^2 c_W^{-2} \right)\sqrt{\fr{2+\beta^2}{(1+\beta^2)(-t_R^2-1)+t_R^2c_W^{-2}}}}, \nonumber \\
m_{33}&&=\frac{4t_R^2\left((w^2+\La_R^2+w_R^2)\left(t_R^2-(1+t_R^2)(1+\beta^2)s_W^2\right)^2+\mathcal{M}_{33}\right)}{3\left(t_R^2-(1+t_R^2)(1+\beta^2)s_W^2+\beta^2 s_W^2 \right)},\nn
\eea
where $\mathcal{M}_{33}$ takes the form,
\bea \mathcal{M}_{33}= && 2t_R^2\beta \left((\sqrt{3}+\beta)\La_R^2+\beta w_R^2\right)s_W^2 +\beta \left( \left(-2\sqrt{3}+\beta-\beta^3 -2t_R^2(\sqrt{3}+\beta) (1+\beta^2)\right)\La_R^2\right)s_W^4\nonumber \\&&  -\beta s_W^4 \left(2+\beta^2+2t_R^2(1+\beta^2)\right)w_R^2.\nn\eea
 

\end{document}